\documentclass[preprint,10pt]{aastex}

\newcommand{\ee}[1]{\mbox{${} \times 10^{#1}$}}
\newcommand{\eten}[1]{\mbox{$10^{#1}$}}

\newcommand{\degree}{\mbox{$^{\circ}$}}

\newcommand{\as}{\mbox{\arcsec}}

\newcommand{\kms}{\mbox{km s$^{-1}$}}
\newcommand\cmv{\mbox{cm$^{-3}$}}

\def\lsim {$\rlap{\raise.4ex\hbox{$<$}}\lower.55ex\hbox{$\sim$}\,$}

\newcommand{\lsun}{\mbox{L$_\odot$}}
\newcommand{\msun}{\mbox{M$_\odot$}}
\newcommand{\ta}{{$T_A^*$}}

\newcommand{\tr}{\mbox{$T_R^*$}}

\newcommand{\lbol}{\mbox{$L_{bol}$}} 

\newcommand{\mean}[1]{\mbox{$\langle#1\rangle$}} 
\newcommand{\mvir}{\mbox{$M_{vir}$}} 
 
\newcommand{\cs}{CS}
\newcommand{\css}{C$^{34}$S}
\newcommand{\ccs}{$^{13}$CS}
\newcommand{\HII}{\mbox{{\rm H}\,{\scriptsize II}}}

\newcommand{\csfifo}{CS $J = 5 \rightarrow 4$}
\newcommand{\cssfifo}{C$^{34}$S $J = 5 \rightarrow 4$}
\newcommand{\ccsfifo}{$^{13}$CS $J = 5 \rightarrow 4$}
\newcommand{\cssesi}{CS $J = 7 \rightarrow 6$}
 
\input{epsf}

\newcommand{\beam}{\mbox{$\theta_{mb}$}}

\newcommand{\rcs}{\mbox{$R_{CS}$}} 
\newcommand{\rn}{\mbox{$R_{n}$}} 
\newcommand{\ar}{\mbox{$(a/b)_{obs}$}} 
\newcommand{\dv}{\mbox{$\Delta v(C^{32}S)$}} 
\newcommand{\dvs}{\mbox{$\Delta v(C^{34}S)$}} 
\newcommand{\dvc}{\mbox{$\Delta v(^{13}CS)$}} 
\newcommand{\mv}{\mbox{$M_{vir}$}} 
\newcommand{\surfden}{\mbox{$\Sigma$}} 
\newcommand{\lcs}{\mbox{$L({\rm CS54})$}} 
\newcommand{\ltom}{\mbox{$L_{bol}/M_{vir}$}} 
\newcommand{\median}{\mbox{$\mu_{1/2}$}} 

\newcommand{\mtotstar}{\mbox{$M_\star(tot)$}}

\slugcomment{\footnotesize {\LaTeX}ed at \number\time\ min., \today}

\begin{document}

  
\title {\bf A \csfifo\ Mapping Survey Towards High-mass Star 
Forming Cores Associated with Water Masers}
\author {Yancy L. Shirley\altaffilmark{1}, Neal J. Evans II, Kaisa E. Young, Claudia Knez, \& Daniel T. Jaffe}
\affil{Department of Astronomy, The University of Texas at Austin,
       1 University Station C1400, Austin, Texas 78712--1083}
\email{yshirley@aoc.nrao.edu}
\email{nje@astro.as.utexas.edu}
\email{kaisa@astro.as.utexas.edu}
\email{claudia@astro.as.utexas.edu}
\email{dtj@astro.as.utexas.edu}
\altaffiltext{1}{Jansky Postdoctoral Fellow, Current Address: NRAO, P.O. Box O, Socorro, NM 87801}

 
\begin{abstract}

We have mapped 63 regions forming high-mass stars in \csfifo\ using
the CSO.  The CS peak position was observed in \cssfifo\ towards 57
cores and in \ccsfifo\ towards the 9 brightest cores.
The sample is a subset of a sample originally selected toward
water masers; the selection on maser sources should favor sources in an
early stage of evolution.
The cores are located in the first and second galactic quadrants
with an average distance of $5.3 \pm 3.7$ kpc and were well detected with
a median peak signal-to-noise in the integrated intensity of $40$.
The integrated intensity of \csfifo\ correlates very well with the
dust continuum emission at 350 \micron.
For 57 sufficiently isolated cores, a well-defined angular 
size (FWHM) was determined. 
The core radius (\rcs), aspect ratio (\ar), 
virial mass (\mv), surface density (\surfden), and the luminosity
in the \csfifo\ line (\lcs) are calculated.
The distributions of size, virial mass, surface density, and luminosity 
are all peaked with a few cores skewed towards much larger values than the
mean. 
The median values, \median , are as follows: 
\median (\rcs ) $= 0.32$ pc, \median (\ar ) $= 1.20$, \median (\mv ) $= 920$ \msun,
\median (\surfden ) $= 0.60$ g cm$^{-2}$, \median (\lcs ) $=1.9 \ee{-2}$ \lsun ,
and \median (\ltom ) $= 165$ \lsun /\msun .
We find a weak correlation between \css\ linewidth and size, consistent with
$\Delta v \sim R^{0.3}$.  The linewidths are much higher than would be predicted
by the usual relations between linewidth and size determined from regions of lower mass.
These regions are very turbulent.  
The derived virial mass agrees within a 
factor of 2 to 3 with mass estimates from dust emission at 350 \micron\
after corrections for the density structure are accounted for.  The resulting
cumulative mass spectrum of cores above 1000 \msun\ can be approximated
by a power law with a slope of about $-0.9$, 
steeper than that of clouds measured with tracers of lower density gas
and close to that for the total masses of stars in OB associations.  
The median turbulent pressures are comparable to those in UC\HII\ regions,
and the pressures at small radii are similar to those in hypercompact
\HII\ regions ($P/k \sim \eten{10}$ K\cmv). The filling factors for dense
gas are substantial, and the median abundance of CS is about \eten{-9}.
The ratio of bolometric luminosity to virial mass is much higher than the 
value found for molecular clouds as a whole, and the correlation of 
luminosity with mass is tighter.

\end{abstract}

\keywords{stars: formation  --- ISM: dust, extinction --- ISM: clouds}


\section{Introduction}

Many, possibly most, stars form in clustered environments with massive
stars (see Carpenter 2000). Regions forming massive stars are the 
only detectable manifestations 
of star formation in other galaxies. Understanding the formation of massive
stars is crucial to an improved understanding of galaxy formation.
Despite all these motivations, our understanding of the conditions in 
which massive stars form is quite primitive. In contrast to the well-developed
theories for isolated, low-mass star formation (e.g., Shu, Adams, \& 
Lizano 1987), 
theories dealing with massive star formation are less developed. While promising
theoretical work has been done (e.g., Bonnell et al. 1997, Bonnell, Bate, \& Zinnecker 1998, 
Klessen 2001, McKee \& Tan 2002, 2003)
the theoreticians are hampered by a lack of systematic information on the
properties of the regions. Many detailed observational studies of individual 
regions have been made, but the field has lacked statistical information based
on large samples analyzed with uniform methods.

One approach to this problem has been to collect a unified data base for
a well-characterized sample. Most work of this kind has focused on samples
selected to have ultra-compact \HII\ regions or {\it IRAS} colors similar to
those of cores with UC\HII\ regions 
(Wood \& Churchwell 1989, Sridharan et al. 2002). The sample studied by
Sridharan et al. and Beuther et al. (2002) used {\it IRAS} colors, but
then selected against \HII\ regions by choosing sources with low emission
in the radio continuum in an attempt to identify early phases.
We have sought to study an early phase by selecting sources based on their
water maser emission (Cesaroni et al. 1988). A survey of a large sample of water
masers revealed that emission in the \cssesi\ transition was common in 
this sample (Plume et al. 1992; hereafter Paper I). 
Detection of this highly excited line suggested high densities
and temperatures, but additional transitions were needed to pin down the
conditions. A multi-transition study of CS lines showed that the 
density, $n$(\cmv), of the sample of 71 sources was characterized by
$\mean{\log n} = 5.9$
(Plume et al. 1997; hereafter Paper II). 
That study also made cross-scans of 25 sources
to estimate sizes, masses, and star formation activities, indicated by
the luminosity to mass ratio ($L_{bol}/M_{vir}$), where the mass 
referred to the dense gas probed by CS.

In the current paper, we present fully sampled maps in the \csfifo\ line
of many more sources (63) than were mapped in Paper II.
These data should provide a much firmer
statistical foundation for determining the conditions at early stages of
the formation of massive stars. 
We have made similar maps of \cssesi\ and dust continuum emission for a subset of these
sources. These data will allow a more detailed analysis of the density and
temperature gradients, similar to that accomplished by van der Tak et al.
(2000) on a small subset of these sources. The analysis of the dust continuum
data (Mueller et al. 2002b) and combined models of CS excitation 
will be presented separately.
While there is a wealth of information on velocity structure in this
data set, we focus on the integrated intensity maps in this paper. 
For an example of interesting velocity structure in the S235 region,
see Lee et al. (2002). 
A summary of early results of this work
can be found in Evans et al. (2002), Shirley et al. (2002), Mueller et al. (2002a),
Knez et al. (2002a), and Lee et al. (2002).

\section{Sources and Observations }

Sixty-three high-mass star forming cores (typically \mvir\ $> 50$ \msun )
were mapped in the
\csfifo\ transition between September 1996 and July 1999 at the 
Caltech Submillimeter Telescope (CSO).
Fifty-seven cores were observed in the \cssfifo\ transition 
and nine cores were observed in the \ccsfifo\ transition towards the
C$^{32}$S peak position between July 2001 and June 2002 at the CSO.
We employ the conventional notation that, unless noted otherwise, the
isotope is the most common one: thus CS means $^{12}$C$^{32}$S.

\subsection{The Sample}

All of the objects observed are listed in Table 1 and Table 2.
Nearly all of the cores
are located in the first and second quadrant (Figure 1). 
Sources were selected from Paper I and Paper II based on
the strength of their \cssesi\ emission with  
each source detected at the 1 K $T_R^*$
level (Paper I). Within this criterion, we made some effort to
include sources with weaker emission. 
This sample extends the sample of 25 cores mapped in
Paper II by including less massive cores and fully mapping each core.
The center of each map was the water maser position from the catalog 
of Cesaroni et al. (1988).

The sources were distributed from $0.7$ kpc to $15.6$ kpc from the sun
(Figure 1).
The distances were determined from an extensive literature search 
(See Table 1, 2 for distance references).  
Photometric distances were used whenever possible, but distance estimates 
to many
cores are based on kinematical distances using the rotation curve of Fich et al.
(1989).  The average distance of the sample of 63 cores is $5.3 \pm 3.7$ kpc
while the median distance is $4.0$ kpc.  The distribution is strongly
peaked between $2$ to $4$ kpc.
The sources at large distances from us are all in the first quadrant.  
The distances can be converted into galactocentric distances, $D_g$,
using a distance of 8.5 kpc to the solar circle.
The result is an average distance of $7.3 \pm 2.6$ kpc 
and a median distance of $6.8$ kpc from the galactic center.
Most (64\%) of the cores are located between $5$ and $10$ kpc from the
galactic center, 25\%\ of the cores are less than $5$ kpc from the galactic
center, and 11\%\ are beyond $10$ kpc (Figure 1). 
This sample is characterized by regions near 
the solar galactocentric distance within the galaxy.

	There is very little overlap of previous CS studies (J$_u \geq 2$) 
selected towards water maser positions: 3 sources in common
(Zinchenko et al. 1994); 0 sources in common (Zinchenko et al. 1995);
0 sources in common (Juvela 1996); and 8 sources in common (Zinchenko et al. 1998). 
There is slightly more overlap of sources selected towards UC\HII\ regions
or IRAS colors indicative of UC\HII\ regions:
24 sources in common (Bronfman et al. 1996);
6 sources in common (Olmi \& Cesaroni 1999); and  3 sources in common (Beuther et al. 2002). 
Thirty-two of our sources were included in the CS $J = 1 \rightarrow 0$  and 
NH$_3$ survey of Anglada et al. (1996)
while six sources were included in the N$_2$H$^+$ $J = 1 \rightarrow 0$
survey of Pirogov et al. (2003); however, we trace a denser gas component 
with the $J = 5 \rightarrow 4$ transition of CS.

\subsection{Observational Method}

	The 230 GHz sidecab receiver with a 50 MHz AOS backend was used 
for all observations (Kooi et al. 1992, Kooi et al. 1998).  
The average velocity resolution was 0.119 \kms. The observing parameters and
conditions are
listed in Table 3.  The standard chopper calibration method was used 
to measure \ta\ (Penzias \& Burrus 1973).  
The beam size (\beam) at 244 GHz was
24\farcs5 for the September 1996 through July 1998 observations.
The secondary edge taper was increased from $-5.2$ dB to $-8.5$ dB 
in August 1998 (Chamberlin, R. priv. comm. 2001; see Kooi, J. 1998) 
resulting in a larger beam of 30\farcs5 at 244 GHz (see Table 3).  
Only five sources were mapped using the larger beam size 
(W49S, W51W, DR21S, W75(OH), and CepA). 

	Observations towards the peak of the \csfifo\ integrated intensity
were made in \cssfifo\ and \ccsfifo .  All of the observations were
made after the secondary edge taper was increased.  The beam size at
$241$ GHz and $231$ GHz was 31\farcs0 and 32\farcs5 respectively.

Determinations of the main beam efficiency, $\eta_{mb}$, were made on planets during
each observing run.  The average $\eta_{mb}$ increased by $20\%$
after the secondary edge taper was increased, excluding the last two
observing sessions (01/02 and 06/02) where mirror alignment problems
decreased the main beam efficiency.  
Average system temperatures ranged from 191 K to 590 K during
the observations.  Pointing was checked every hour using planets.
The average standard deviations in azimuth and zenith angle pointing were 
5\as\ and 4\as\ respectively for all of the observations, resulting in a
6\as\ pointing uncertainty.
These errors, adding to about one-quarter beam, are upper limits to the
actual pointing errors because they were mostly slow drifts over the
time of a run and pointing was corrected by repeated measurements
during each night.

	The cores were mapped using the On-The-Fly (OTF) mapping technique 
(e.g., Mangum et al. 2000) with an oversampled 10\as\ grid in RA-DEC coordinates.  
The scan rate was set at $2\as$ per second to provide 5 s 
of integration time per spectrum. On some occasions, the maps were 
repeated for higher signal-to-noise.  The map was extended until 
the \csfifo\ line was not detected or negligible compared to the peak. 
The average rms per spectrum in the maps varied between 0.1 K and 0.6 K.

\section{Results}

The integrated intensity of the \csfifo\ transition was
calculated using,
\begin{eqnarray}
I(T_A^*) & = & \int_{v_1}^{v_2} T_A^* dv \\ 
\sigma _{I(T_A^*)}^2 & = &
\left< \sqrt{\Delta v_{line} \delta v_{chan}} \sigma_{T_A^*} 
\right>^2_{map}  + 
\left( \frac{\Delta v_{line}}{\Delta v_{lft} + \Delta v_{rt}} \right)^2
\sigma_{I_{base}}^2 
\end{eqnarray}
where $\Delta v_{line} = v_2 - v_1$ is a velocity interval that includes the entire
line (as distinct from the FWHM of the line), 
$\Delta v_{lft}$ and $\Delta v_{rt}$ are the velocity intervals
of the left and right baselines, $\delta v_{chan}$ is the spectrometer
velocity resolution, and $\sigma_{I_{base}}$ is
the standard deviation of the integrated intensity of the 
total baseline ($\Delta v_{lft} + \Delta v_{rt}$) calculated
over all of the spectra in the map.  
The first term in the integrated intensity error is the theoretical
error and assumes no deviation from a linear baseline.
The second term in the integrated
intensity error compensates for residual variations in
the baseline after a linear baseline was removed.  This average error
in the integrated intensity is added in quadrature to the average of the
theoretical error for the integrated intensity, calculated for each
spectrum in the map.  The theoretical error (first term) typically
dominates.  The integrated intensity is placed on the \tr\ scale (Kutner \& Ulich 1981)
by dividing Equation (1) by the $\eta _{mb}$ appropriate
for the night the object was observed (Table 3).  This calibration procedure
is described in \S2 of Paper II.  
An assumed error in $\eta _{mb}$ ($10\%$) was propagated
into the uncertainty in $I(T_R^*)$.

Contour maps of integrated intensity are shown in Figures 2 through 12.
The average extent of the maps is $\pm 50$\as , but larger maps 
were made where necessary.
The lowest contour is at least 2$\sigma_I$ and typical contour
intervals are $10\%$ of the peak intensity.  
The cores were well detected with a median peak-signal-to-noise of $40$
and peak integrated intensities that range from $5.5$ to $208$ K \kms.
The median separation of the peak of CS integrated intensity 
from the 350 \micron\ dust continuum peak (Mueller et al. 2002)
is 7\as. 
The peak integrated intensity correlates well ($r =0.85$) with the 
submillimeter
flux at 350 \micron\ (Figure 13);
a fit to the logarithms indicates a relationship that is nearly linear:
$\log I(T_R^*)$ $= (-0.60 \pm 0.01) + (0.92 \pm 0.05)\log S_{350 \micron}$.
Objects that are bright at 350 \micron\ are also strong emitters in the
\csfifo\ line and the dust continuum and CS emission 
are coincident.  Since the 350 \micron\ dust continuum is optically thin, it
is a good tracer of mass along each line-of-sight.  
The strong correlation between
$I(T_R^*)$ and $S_{350 \micron}$ confirms that high J lines of CS are 
excellent tracers of dense, warm gas.

An extensive literature search using the SIMBAD database was performed
to find \HII\ regions associated with the dense CS cores.  Only twelve cores
($19\% $) were found with no obvious, direct association with 
radio continuum emission (e.g., UC\HII ).
When possible, the 2 cm size is reported is Tables 1 \& 2.  Using the
taxonomy of Kurtz (2002), \HII\ regions are classified as
ultra-compact (UC\HII ) if the diameter is $\leq 0.1$ pc, compact
(C\HII ) if the diameter is $\leq 0.5$ pc, and an extended \HII\ region
if the diameter is greater than $0.5$ pc or clearly associated with a 
classical 
\HII\ region.  The classifications of several UC\HII\ are unclear
since no 2 cm sizes are reported in the literature. 
Some of the cores contain multiple UC\HII\ regions (see Conti \& Blum 2002, W49N region),
but only the nearest \HII\ region to the water maser peak 
is plotted to minimize obscuration of the CS map (Figures 2 - 12). 
The CS integrated intensity is weaker for sources with no known radio
continuum emission 
($\mean{I(\tr )} = 25.9 \pm 23.1$ K \kms , 
$\mu_{1/2} = 19.7$ K \kms ) than for sources with UC\HII\ regions
($\mean{I(\tr )} = 50.4 \pm 41.0$ K \kms , $\mu_{1/2} = 35.7$ K \kms )
and C\HII\ or \HII\ regions 
($\mean{I(\tr )} = 54.6 \pm 58.0$ K \kms , $\mu_{1/2} = 32.0$ K \kms ).

The \cs\ centroid is generally close to the water maser peak with
a median centroid distance of $8$\as ; 
only eight cores ($13\% $)
have \cs\ centroids more than $\theta_{mb}/2$ away from the water maser
position.     
The median distance between the \cs\ centroid and 
\HII\ regions was 8\as , less than
one third of the beam FWHM but larger than the average pointing
uncertainty.  The peak CS
emission is directly associated with the \HII\ region in thirty-six ($57\%$)
of those cores while fifteen \HII\ regions are more than $\theta _{mb}/2$
away from the CS peak.  The dense gas traced by \csfifo\ emission is
clearly associated with water maser emission and often
associated with an (UC,C-)\HII\ region.

	The majority of cores ($46$) are isolated within the regions 
mapped ($1.7$\arcmin\ field-of-view for the average map size).
Seventeen cores ($27\%$) have companions with a  
median separation of $0.93$ pc.
Three cores have more than 2 distinct companions within the mapped region 
(S87, W51W, W75(OH)).

	Spectra towards the W49N region (also denoted W49A North)
display two blended velocity components.  The \csfifo\ lines
clearly show a peak near $4$ \kms\ and $12$ \kms\ in all spectra in the map.
There is considerable debate
in the literature over the correct interpretation of the two velocity
components: are there multiple clouds (see Serabyn et al. 1993) 
or is this purely an optical depth effect (see Dickel et al. 1999)?
Since the two components are also observed in the \css\ and \ccs\ isotopomers,
we shall analyze W49N as two separate clouds with the caveat that
this region is very complicated.  The integrated intensity for two-component
Gaussian fits to the spectra are shown in Figure 8.

	The integrated intensities for \cssfifo\ and \ccsfifo\ observations
are listed in Tables 4 and 5.  Forty-nine cores were detected in the \cssfifo\
transition while seven cores were not detected
to an average $3 \sigma$  \ta\  level of $300$ mK (G135.28+2.80, S241, S252A, 
G24.49$-$0.04, S106, BFS11$-$B, S157).  The average integrated intensity 
is $\mean{I(\tr )} = 6.5 \pm 7.5$ K \kms\
with a median of $4.0$ K \kms, both values a
factor of $10$ lower than the corresponding values for \csfifo .
Nine of the strongest cores were also observed in \ccsfifo\ with all of the
cores detected.  The average ratio between the integrated intensity of
\cssfifo\ and \ccsfifo\ is $2.6$, consistent with the observed interstellar 
isotope ratio between $^{34}$S and $^{13}$C (Wilson \& Rood 1994).

\section{Analysis}

\subsection{Core Size \& Aspect Ratio}

Previous studies (e.g., van der Tak et al. 2000, Hatchell et al. 2000,
Beuther et al. 2002) and our modeling of the dust continuum emission (Mueller et al. 2002b)
indicate that the distribution of density is well fitted by a power law,
$n(r) \propto r^{-p}$.
Since power laws have no intrinsic size scale, assigning a size to such
distributions can be highly misleading. Following long tradition, we will
calculate a nominal radius for each source from a Gaussian deconvolution of
the beam, and we will use this radius for calculation of masses. 
We caution that this radius should be viewed strictly as a fiducial
radius, with no physical significance. We discuss later the likely corrections
to masses, etc. that result from continuation of power laws to larger scales.

The angular extent of each map at the half power level
was determined by finding the area 
within the contour at half $I_{peak}$, $A_{1/2}$, and calculating the angular
radius of a circle with the same area.  
The nominal core radius, \rcs, was determined by
deconvolving the telescope beam (\beam ) assuming both are Gaussians: 
\begin{equation}
\rcs = D \left( \frac{A_{1/2}}{\pi} - \frac{\theta _{mb}^2}{4} \right)^{1/2} \; ,
\end{equation}
where $D$ is the distance to the core.  Similarly, the deconvolved angular
size, $\theta _{dec}$, is found from
\begin{equation}
\theta _{dec} = \left( \frac{4 A_{1/2}}{\pi} - \theta _{mb}^2 \right)^{1/2} \; .
\end{equation}
The core radius and uncertainty
are listed in Tables 6 and 7.  The quoted uncertainty in core radius 
is derived from the uncertainty in area of the core ($A_{1/2}$)
and the uncertainty in the main beam FWHM, assumed to be 10\%\ of $\theta _{mb}$.  
The distance uncertainty actually dominates the uncertainty in \rcs, 
but it is ignored in this analysis since
$\sigma _D$ is difficult to determine.  Since the distance may be uncertain
by at least 50\%, the core radius would be uncertain by the same factor.

Almost all (57) of the cores have clearly defined values for \rcs.  
The remainder (6 cores) have multiple peaks too close together to allow
unambiguous determination of a FWHM angular size (see Tables 6 and 7).
This sample of 57 cores provides the sample for the statistical 
analysis in the rest of the paper.
The majority of cores (36, $63\% $) have {\it deconvolved} sizes that are 
larger than the main beam FWHM, indicating that they are
well resolved (Figure 14).  
The dashed line in Figure 14 indicates the \rcs\ at 
each distance for which the deconvolved source size equals the FWHM beamsize.
The largest core was W49S with $\rcs = 1.53$ pc,
while the smallest cores were S252A and G121.30+0.66, with $\rcs = 0.10$ pc.

The average over the sample is $\rcs = 0.37 \pm 0.26$ pc
while the median core size is $0.32$ pc.
The distribution of $\log \rcs$ is peaked for core sizes
near the mean and median values (Figure 15a). 
For a source at the median distance of the sample, $4.0$ kpc,
$\rcs < 0.19$ pc would fail our criterion ($\theta_{dec} \ge \theta_{mb}$)
for being well-resolved.
The median distance bias is shown as a horizontal dotted line in Figure 14.  
The average over the sample is 
smaller than the average core radius of $0.5 \pm 0.4$ pc 
determined in Paper II for the 25 cores with cross-scans. 
Sources not directly associated with radio continuum emission (N $= 12$)
are slightly smaller than cores with radio continuum emission 
($\rcs = 0.28 \pm 0.14$ pc and $\mu_{1/2} = 0.25$ pc; see Figure 15a).
There is no statistically significant difference between the sizes of
cores associated with UC\HII , C\HII , or \HII\ regions and the
complete sample.

The process of finding a FWHM size might vary with the intensity of the
core, introducing a bias into the size distribution.  The integrated
intensity is plotted against \rcs\ in Figure 14 for 57 cores.  There is no 
observed correlation ($r = 0.07$) between CS intensity and core radius
over a wide range in both variables.

The 350 \micron\ dust continuum from twenty-four sources from our survey were modeled 
with a radiative transfer code by Mueller et al. (2002b). 
The best fit power law index, $p = - \log n/\log r$, is listed in 
Tables 8 and 9.  Convolving a 
power law intensity distribution with a Gaussian beam pattern should
result in deconvolved core sizes that are somewhat larger than
$\theta_{mb}$ (e.g. Terebey et al 1993).  Flatter power laws 
produce larger deconvolved source sizes than steeper power laws.
This correlation was observed towards a sample of low mass cores
observed at 850 \micron\ with SCUBA (Shirley et al. 2000, Young et al. 2003). 
A weak correlation ($r =-0.55$) is observed between the best fit power law index 
and the deconvolved source size determined from our CS maps (Figure 16). 
This correlation is likely real since the observations 
were made with two different instruments, SHARC (Hunter
et al. 1996) and the CSO 230 GHz receiver, with different beam sizes ($14$\as\ and
$24\farcs 5$ respectively).  For power law density distributions, the deconvolved
source size may be used as a rough guess of $p$ if the correlation is
calibrated.

An alternative method for determining the core radius is to measure the
sizes of the cores at the same intensity level.  For instance, the
core radius, $R_{10}$, at an intensity level of $I(\tr ) = 10$ K km/s
is calculated (Tables 6 \& 7) using the
same method as for \rcs\ and deconvolving a Gaussian telescope beam:
\begin{equation}
R_{10} = D \left[ \frac{A_{10}}{\pi} - \frac{\theta _{mb}^2\ln (I_{\rm{peak}}(\tr )/10)}{4 \ln 2} \right]^{1/2} \; .
\end{equation}
$R_{10}$ was unresolved for 11 cores ($R_{10} <$ \rcs\ ), was too large to be
determined for 6 cores ($R_{10} >$ extent of the map), and encompassed more
than one core in 4 cases.  The average core size for 33 cores 
was $0.50 \pm 0.32$ pc with a median size of $0.43$, 35\% larger than for
\rcs. Since the choice of the intensity level is arbitrary and \rcs\
can be defined for many more cores, \rcs\ is
the core radius used in most comparisons and calculations in this paper.
$R_{10}$ is used in \S4.2 for an alternative
calculation of the linewidth-size relationship to explore the sensitivity
of the results to this definition.

Aspect ratios for each core were determined from the ratio of major to
minor axis for the $20\%$ peak contour (Tables 6 and 7).  The $20\%$ peak
contour is well detected and resolved for the entire sample 
($\mean{\sigma_{20\%}} = 10\sigma_I$).  
The distribution of aspect ratios (Figure 15b) is strongly peaked 
towards low (\ar\ $< 1.4$) aspect ratios indicating that 
the observed contours are consistent with circular symmetry. 
The mean aspect ratio is $1.26 \pm 0.22$ while the 
largest observed aspect ratio is $1.8$
(ON2S).  The cores are observed in projection, making
\ar\ a lower limit to the actual aspect ratio. 
The position angle of the major axis, measured counter-clockwise from north, is 
listed in Tables 6 and 7.  The histogram of position angles for cores
with \ar\ $\geq 1.2$ is plotted in Figure 15c.  
There is no bias in the core elongation 
observed along the scan direction of the OTF map ($90\degree $), indicating
that the aspect ratios are unaffected by any beam smearing from the OTF method.

Young et al. (2002) report a correlation
between \ar\ and $p$ toward low-mass cores; flatter
power laws ($p \sim 1$) are associated with more elongated cores.
Using the $p$ values from Mueller et al. (2002) and the CS aspect ratios,
we find no evidence for a correlation in this sample (Figure 16).  
It is necessary to use the CS data to determine the 
aspect ratio since Mueller et al. were unable to determine reliable
aspect ratios because of the effects of chopping.

\subsection{Linewidth-Size Relationship}

The FWHM linewidth, \dv, for each core was determined from a 
Gaussian fit to a spectrum produced by convolving the data to
an effective size corresponding to the half-power contour.  
The average linewidth for the sample of 63 cores was
$\dv = 5.6 \pm 2.2$ \kms.  A few cores show evidence for self-absorption
and other optical depth effects.
For a Gaussian line shape, the broadening due to
optical depth can be expressed by
\begin{equation}
\frac{\Delta v}{\Delta v_o} = \frac{1}{\sqrt{\ln 2}} \;
\sqrt{ \ln{  \frac{\tau }{\ln{ \frac{2}{1 + \exp (-\tau )}
                  }  }
                  } } \; ,
\end{equation}
where $\Delta v_o$ is the optically thin linewidth
(Phillips et al. 1979).  We can use the \cssfifo\ linewidth
to test optical depth effects.  The \cs\ to \css\ linewidth ratio for
49 cores was $1.3 \pm 0.4$ corresponding to an average optical 
depth of $\tau = 1.7$ (Figure 17).  Therefore, \dvs\ 
should be used when possible in calculations 
sensitive to the linewidth.  We checked the optical depth of
\css\ by observing \ccsfifo\ towards nine of the brightest cores.
The \css\ linewidths are consistent with being optically thin
for all but 3 cores (Figure 17).

The linewidth-size relationship for 51 cores using \dvs\ 
is plotted in Figure 18.
The data were fitted with a least-squares method, including statistical 
errors in both quantities (Press et al. 1992) to give
${\rm log}\dvs = (0.92 \pm 0.01) +  (0.43 \pm 0.02) \log \rcs$.
The linear correlation coefficient is low ($r = 0.36$).
For comparison, a fit using robust estimation (Press et al. 1992),
which is less sensitive to outliers, 
gives a shallower slope,
${\rm log}\dvs = 0.77 +  (0.17) \log \rcs$.
If we average these two slopes, then \dvs\ is 
roughly proportional to $\rcs ^{0.3}$.
This slope of the linewidth-size relationship is consistent with the findings
of Caselli \& Myers (1995), who find a shallower linewidth-size
relationship for ``high-mass'' regions ($\Delta v \propto R^{0.21\pm 0.03}$) 
compared to `low-mass'' ($<$ few \msun ) regions ($\Delta v \propto R^{0.53\pm 0.07}$)
probed by $^{13}$CO and C$^{18}$O.  Various studies
using different tracers (NH$_3$, $^{12}$CO, etc.) find linewidth-size
relationships that vary between $\Delta v \sim R^{0.2}$ to $R^{0.8}$.
(e.g., Brand \& Wouterloot 1995; Jijina, Myers, \& Adams 1999; 
and Brand et al. 2001).  Alternatively, if we calculate the linewidth-size
relationship using $R_{10}$ instead of \rcs\ (Figure 18, bottom
panel), the least squares fit ($r = 0.43$), 
${\rm log}\dvs = (0.87 \pm 0.01) +  (0.65 \pm 0.03) \log R_{10}$,
and robust estimation,
${\rm log}\dvs = 0.78 +  (0.20) \log R_{10}$, do not agree well.
It is difficult to rigorously compare 
the results from our sample because the correlations are very weak
and the distance uncertainty is large enough to eliminate the observed
correlations.

The more important point is that the linewidths are all
much larger at a given radius than those found in either ``low-mass''
or ``high-mass'' regions by Caselli \& Myers (1995).  
For the average
core size in our sample, the average \css\ linewidth is $4$ times
larger than the ``high-mass'' prediction and 5 times larger than
the ``low-mass'' prediction of Caselli \& Myers (1995).
This point, already made in PaperII, is strengthened by
the larger sample and fully sampled maps presented here.
Note that the ``high-mass" regions of Caselli et al.,
observed towards Orion, cover a similar range of radii but are 
less massive than those studied here.  
Extension of the linewidth-size relation found in previous studies
to regions of massive star formation would be very misleading.

We attribute the large linewidths to turbulent motions since
the thermal contribution to the linewidth is negligible.  Assuming $T_k = 50$ K, 
thermal broadening accounts for only $0.23$ \kms, whereas 
the smallest linewidth in our sample is $2.49$ \kms\ (S87).  The sonic linewidth
for $T_k = 50$ K gas  and a mean molecular weight of $\mu = 2.29$ is $1.0$ \kms .
Outflows are
apparent in line wings for some sources, but they are unlikely to broaden
the FWHM linewidth, except by stirring up turbulence.
The turbulent linewidth of this sample is highly supersonic.
Our regions are at least four times more turbulent than regions involved 
in lower mass star formation (see Mardones et al. 1997, Gregersen et al. 1997). 
Based on comparison of power-law models
using dust emission, Mueller et al. (2002) found that these cores were also
about 100 times denser on average than the low mass sample.

\subsection{Virial Mass}

The virial mass for a homoeoidal ellipsoid (concentric 
ellipsoids of revolution with equal aspect ratios) is given by
\begin{eqnarray}
\mv (R) & = & \frac{5 R \Delta v^2}
                  {8 a_1 a_2 G \ln 2 }  \;\; 
          \approx 209 \; \frac{ \left( R / 1 \rm{pc} \right) 
           \left( \Delta v / 1 \rm{km s^{-1}}  
\right)^2 }{a_1 a_2} \; \msun \;\; ; \\
a_1 & = & \frac{1 - p/3}{1 - 2p/5} \;\; , \;\; p < 2.5 \;\; 
\end{eqnarray}
where $a_1$ is the correction for a power law 
density distribution and $a_2$ 
is the correction for a non-spherical shape (Bertoldi \& McKee 1992).
For aspect ratios less than $2$, $a_2 \sim 1$ and can be
ignored for our sample.
The equation in Bertoldi and McKee uses an rms velocity; we have converted
to the observable ($\Delta v$) under the assumption that turbulent broadening
dominates thermal broadening; this is a very safe assumption for these
sources, but it fails for lines of light species in very quiescent
regions (see Shirley et al. 2002b).  

There are several corrections used in calculating the virial mass.
Since the \cs\ linewidth was found to be optically thick in
some cores, we use the \css\ linewidth when it was observed. 
The remaining cores (7) are corrected using the average ratio of
\css\ to \cs\ linewidth for the sample (\S 4.2). 
We use the density power law index, $p$, from Mueller et al.
(2002) for the cores common to each sample (21) and use the average $p = 1.77$
for the remaining cores.  Finally we must choose a radius within which
to calculate the virial mass.  Initially we use \rcs . However, since a
power law density distribution has no characteristic size, we also
calculate virial masses using \rn , the radius at which the density 
of the dust models drops to $10^4$ \cmv\ (Mueller et al. 2002b). This density 
corresponds to the density of the ambient molecular cloud at the edge of a core 
based on a detailed study of molecular clouds in our galaxy (Allers et al. 2003).
The average \rn\ $= 0.40$ pc is only slightly larger than the average \rcs .
The virial mass using corrections for 
$\mean{p}$ and $\mean{\Delta v(C^{34}S)/\Delta v(C^{32}S)}$ 
is $2.3$ times smaller than the mass calculated using
\dv\ and assuming a constant density envelope.  It is crucial to
account for variations in density structure and optical depth
effects when calculating the virial mass.

The distributions of virial masses are peaked near $1000$ \msun, for either
definition of the cloud radius (Figures 15d \& 15e).  
Only cores for which all the corrections could be made ($21$)
are included in the $M_{vir}(R_n)$ histogram.  Our sample begins to
be incomplete for cores with masses less than about $10^3$ \msun\ because they
will tend to be too small to resolve at the average distance of sources in
our sample ($5.3$ kpc). Consequently, the peaked histograms and the average
values given below should be taken only as representative of this
particular sample.  A weak correlation is observed between 
$\log \mv$ and $\log I(T_R^*)$ ($r = 0.43$), indicating that more massive
cores are typically brighter in CS emission (Figure 19).

The median virial mass using \rcs\ is $920$ \msun\ 
for the full sample of $57$ cores.  
The large dispersion about the mean mass ($\pm 2810$ \msun, see Table 10) 
partially results from the mass
of W49N, which is $8$ times higher than the average mass; therefore, the
median is a better indicator of the typical virial mass for this sample. 
Source without known radio continuum emission (N $= 12$) have a smaller
median virial mass of $329$ \msun .  As was found for the
size distributions, there is no statistically significant
difference between the median virial masses of sources with UC\HII , C\HII ,
or \HII\ regions and the complete sample.
The median virial mass using \rn\ is $610$ \msun\ using the subsample 
of $21$ cores that were
modeled by Mueller et al (2002).

The virial mass may be compared to the mass derived from models
of the dust continuum emission at 350 \micron, denoted $M_{dust}(\rcs )$ 
(Mueller et al. 2002b), for the sources in common.  The average 
ratio of virial mass to dust-determined mass ($\mean{M_{vir}/M_{dust}}$)
is $3.4 \pm 3.3$ and the median ratio is $2.2$ 
for 21 sources with virial mass corrections,
\css\ linewidths, and dust models (see Figure 19).  Given the many sources
of uncertainty in deriving virial and dust-determined masses
(distance, dust opacity, etc.), the agreement is good.
The agreement suggests that the assumptions
used in deriving the virial mass and the choice of Ossenkopf \& Henning
(OH5, 1994) opacities for the dust are sensible, and that virial masses
provide a good mass estimate.

Since the regions we are studying are forming massive stars, we
can compare the virial mass to regions that have formed high-mass stars,
namely OB associations.  Matzner (2002) calculated a mean mass per
association of $440$ \msun\ based on the Galactic H II region luminosity
function of McKee \& Williams (1997).  This mass is roughly $50\%$ of 
the median virial mass calculated using \rcs\ and $75\%$ of the
virial mass calculated using \rn .  If the regions traced by
water maser emission and \csfifo\ emission are forming a single new OB
association, then the star formation efficiency of the gas traced
by high-J CS emission is high ($\sim 50\%$).  However, this star formation
efficiency is an upper limit if more than one dense core contributes
to the formation of a new OB association.

\subsection{The Mass Spectrum}

Because all these cores have masses greater than those of individual stars, 
they are destined to form clusters or associations. 
The cumulative mass spectrum of dense cores should then
be directly related to the cumulative distribution 
of the {\it total} mass of stars in clusters or OB associations (\mtotstar). 
Using the model of McKee and Williams (1997), the cumulative distribution of \mtotstar\
in OB associations is proportional to $\mtotstar^{-1}$.
The mass function of our cores may be related less directly to the 
initial mass function of stars {\it within} those clusters and associations 
(the usual IMF). 
Stars above about 5 \msun\ roughly
follow a power-law mass spectrum ($N(>M) \propto M^{\Gamma}$), with
$\Gamma$ often assumed to be $-1.35$ (Salpeter 1955). Massey et al. (1995)
find $\Gamma = -1.1\pm 0.1$(standard deviation of the mean) for 13 
OB associations. In contrast to these slopes, molecular clouds
as a whole have a flatter distribution.
Mass spectra with $\Gamma$ of $-0.6$ to $-0.7$ have been observed for 
molecular clouds (see Scoville \& Sanders 1987), as well as 
the large clumps within clouds (Blitz 1993, Williams et al. 2000, Kramer at al.
1998).  Studies of cores forming low-mass stars in Ophiuchus reveal a steeper 
mass spectrum, $\Gamma = -1.5$ (Motte, Andr\'e, \& Neri 1998,
Johnstone et al. 2000), and a study in Serpens finds $\Gamma = -1.1$
(Testi \& Sargent 1998).  These slopes begin to resemble the slope of the
the IMF for massive stars, but they mostly apply to lower mass regions where the
stellar IMF actually turns over (Scalo 1998, Meyer et al. 2000).

The cumulative mass spectrum of cores, based on the corrected virial masses,
is shown in Figure 20. The mass spectrum is clearly incomplete below
about $1000$ \msun. The spectrum for $M_{vir} \geq 1000$ \msun\ was fitted
using least squares and robust estimation (Figure 20), with resulting
$\Gamma = -0.91 \pm 0.17$ and $\Gamma = -0.95$, respectively. 
The mass function of dense cores is similar to that of \mtotstar\ in the model
of McKee and Williams (1997). It is also within the range of the values
for the IMF of stars within OB associations (Massey et al. 1995).
The similarity of our value for $\Gamma$ to that of the IMF of stars
within clusters suggests that the fragmentation process keeps
nearly the same mass spectrum.

Our mass spectrum is slightly steeper than found by other studies towards 
high-mass star forming regions that used probes that trace lower densities.  
Kramer et al. (1998) find 
$\Gamma = -0.6$ to $-0.8$ for CO clumps within seven high-mass star forming
clouds. A CS $J = 2 \rightarrow 1$ survey towards fifty-five 
dense cores containing water masers found $\Gamma = -0.6 \pm 0.3$ 
(Zinchencko et al. 1998).

\subsection{Surface Density, Pressure, and Confinement of UC\HII\ Regions}

McKee and Tan (2002, 2003) have emphasized the importance of the surface density
of a molecular core (which they call a clump) in the stellar mass accretion rate 
($dm_*/dt \propto \surfden^{0.75}$)
and the time to form a star
($t_{*f} \propto \surfden^{-0.75}$). Based on the results in Paper II,
they assumed $\surfden = 1.0$ gm cm$^{-2}$.

The surface density of the core can be calculated from
\begin{equation}
\surfden = \frac{\mv (\rcs )}{\pi \rcs^2} \;\; 
         \approx 0.665 \; \frac{ \left( M_{vir}/1.0 \times 10^4 \msun \right) }
                         { \left( \rcs /1 \rm{pc} \right) ^2 } \; 
\rm{gm}\;\rm{cm}^{-2} \;\; .
\end{equation}
The average over the sample with well-determined sizes is
$\surfden = 0.82 \pm 0.78$ gm cm$^{-2}$ with a median of $0.60$ gm cm$^{-2}$.
The median surface density corresponds to $2870$ \msun\ pc$^{-2}$.
The surface densities range from $0.07$ gm cm$^{-2}$ (G58.78+0.06) to 
4.6 gm cm$^{-2}$ (G20.08$-$0.13).  While the distribution is sharply peaked
for \surfden\ $< 1$ gm cm$^{-2}$, a few cores ($6$) have surface densities
greater than $2$ gm cm$^{-2}$ (Figure 15f).
The median surface density would imply a decrease in the mass accretion
rate and increase in the star formation time for the accretion models
of McKee \& Tan (2002, 2003) by a factor of 2/3.  
The picture of McKee \& Tan would imply that cores with higher $\Sigma$ 
should have a higher star formation rate.  Then one might expect
the luminosity to correlate with $\Sigma$. We see no correlation (r $= -0.06$)
in our data, but the range of $\Sigma$ is small. 

The large surface densities and linewidths also translate into high pressures,
both thermal and turbulent. Using equation A6 from McKee \& Tan (2003),
\begin{equation}
\mean{\bar P/k} \approx 4.25 \times 10^8 \; (\surfden / 1 \rm{gm}\, \rm{cm}^{-2})^2 \; K \; \cmv \;\; ,
\end{equation}
we compute a mean pressure for each of our cores with a known surface density.
The average over these cores is $\mean{\bar P/k} = (5.4\pm12.6) \times 10^8$  K\cmv,
with a median value of 1.5\ee{8} K\cmv. The distribution is highly skewed by 
the core with very high surface density, so the median is more representative.

These high pressures may have some bearing on the issue of confinement of
UC\HII\ regions (see De Pree, Rodriguez, \& Goss 1995). 
Simple considerations suggest that the thermal pressure
of an UC\HII\ with $T_e = \eten4$ K and $n_e = \eten4$ \cmv\ could be balanced
by the median pressure in these cores. 
The pressure would be even higher close to the center of the cores. 
Mueller et al. (2002) found a median density over $1.4\ee7$ \cmv\ 
at the fiducial radius of 1000 AU, and a median temperature of 260 K,
leading to a thermal pressure of 4\ee9 K\cmv. Including turbulent
pressure raises this to about 1.5\ee{10} K\cmv, comparable to
those in the newly discovered hypercompact \HII\ regions,
which have sizes on the order of 1000 AU (e.g., Kurtz \& Franco 2002).

While the issues surrounding UC\HII\ regions
are complicated (see Kurtz et al. 2000 for a review), our data do generally 
agree with the idea that turbulent pressure in the surrounding molecular
gas may affect the evolution of \HII\ regions (Xie et al. 1996).
Xie et al. have suggested an anticorrelation between the turbulent
linewidth and the size of an UC\HII\ region for a sample of eight sources.
We do not find strong evidence for an anticorrelation ($r = -0.29$)
between \dvs\ ($r = -0.12$) or $\mean{\bar P/k}$ ($r = -0.29$) and 
UC\HII\ region sizes (Tables 1 and 2); however, the linewidth and mean
pressure determined from \csfifo\ observations with a large beamsize
is probably not the best tracer of the gas that may be directly associated
with confinement of the UC\HII\ region.

\subsection{Filling Factor and CS abundance}

The constant density volume filling factor was calculated by taking the ratio of
the constant density virial mass ($p = 0$) to the mass calculated from the volume density.
The volume density was taken to be the best fit density from the LVG
models, $n_{lvg}$, using multiple transitions of \cs\ and \css\ (Paper II),
\begin{equation}
f_{\rm{v}}(p = 0) = \frac{M_{vir}(\rcs ;p=0)}{\frac{4}{3}\pi \mu m_{H} n_{lvg} \rcs ^3} \;\;
           \approx 0.042 \; \frac{ \left( M_{vir} /
                                 1.0\times 10^4 \msun \right) }
             { \left( n_{lvg} / 1.0\times 10^6 \cmv \right)
               \left( \rcs / 1 \rm{pc} \right) ^3 } \;\;  .
\end{equation}

The average filling factor is $0.46 \pm 0.72$ with a
median of $0.13$ for the subsample of 42 cores for which 
$n_{lvg}$ was determined.  
Paper II found an average filling factor of $0.33 \pm 0.59$,
consistent with the mean of our sample. However, $f_{\rm{v}}(p = 0)$
underestimates the filling factor when there is a density gradient.  
The LVG models of Paper II assume a constant density envelope; 
therefore, $n_{lvg}$ represents an average density that is strongly
weighted toward the denser gas.  
Using the power law models of Mueller et al (2002), 
the mean $n_{lvg}$ corresponds to the density at a 
radius of $7300 \pm 5200$ AU or about 0.1 times the average \rcs .  
Detailed models of sources will allow us to determine $f_{\rm{v}}$ more
accurately, but this comparison suggests that the average core is
not highly clumped in the sense of being mostly empty space with a
small volume filling factor of very dense clumps probed by the CS emission.

In a similar way, we can compare the mass calculated from the CS column
density to the virial mass to constrain the CS abundance, 
\begin{equation}
X(CS) = \frac{\mu m_H N_{lvg} \pi \rcs ^2}{M_{vir}(\rcs )} \;\;
\approx 5.75 \times 10^{-10} \; \frac{ \left( N_{lvg} / 1.0\times 10^{14}
\rm{cm}^{-2} \right)
           \left( \rcs / 1 \rm{pc} \right) ^2 }
         { \left( M_{vir} / 1.0\times 10^4 \msun \right) } \;\; .
\end{equation}
The column density was determined from the LVG models of multiple transitions of
CS lines (Paper II).  The resulting
median value of $X$(CS) is $1.1\ee{-9}$, with a distribution 
(Figure 15g) highly skewed by  large abundances 
in G10.6$-$0.4 ($X(CS) = 1.3\ee{-8}$), in W28A2(1) ($X(CS) = 1.6\ee{-8}$), and 
in W51 ($X(CS) = 3.6\ee{-8}$). The mean and median 
abundances are three times higher than those
found in Paper II.

\subsection{Luminosity of CS}

	The luminosity of \csfifo\ emission was calculated from
\begin{equation}
\lcs \approx  1.05\times 10^{-5} \; \left( \frac{D}{1 \rm{kpc}} \right)^2 
              \left( \frac{\Omega_{source} + \Omega_{beam}}{\Omega_{beam}} \right)
	      \left( \frac{\int T_R^* dv}{1 \rm{K \kms}} \right) \; \lsun \;\; ,
\end{equation}
using the deconvolved source size and assuming that the
source is described by a Gaussian brightness distribution (Paper II).  The
average CS(5--4) luminosity is $(5.0 \pm 8.8) \times 10^{-2}$ \lsun\ for the 
sample of 57 cores, similar to the average CS(5--4) luminosity 
from Paper II ($4.0 \times 10^{-2}$).  The distribution of CS luminosities
is strongly peaked with a tail of high luminosity sources (Figure 15h).  
The median \lcs\ is $1.9 \times 10^{-2}$ \lsun , lower than the average luminosity from 
Paper II, because our sample has included more of the less luminous cores.
The total \lcs\ for our subsample of 57 cores is $2.85$ \lsun .

By estimating the number of star forming cores emitting \csfifo\ we can estimate
the total galactic \lcs .
The latest update to the Arcetri H$_2$O maser catalog (Valdettaro et al. 2001)
indicates 410 regions with H$_2$O masers 
that have IRAS colors indicative of star formation.
Paper II had a detection rate of $75\%$ towards a subset of that sample.
Also correcting for the unobserved portion of the Galaxy in the Arcetri
survey, roughly 1/3 of the sky, we find that there are roughly $460$ cores
detectable in the \csfifo\ line in our Galaxy.  This number may be an
understimate since water masers are variable and some water maser
sources may not have been detected in the Arcetri catalog.
We add the total
luminosity from our subsample to the mean \lcs\ for the remaining \csfifo\
emitting clouds ($460 - 57 = 403$) to find a galactic luminosity, $L_{gal}({\rm CS54})$, of 
$23$ \lsun .  If the median \lcs\ is used, $L_{gal}({\rm CS54}) \approx 11$ \lsun .
If the detection rate is $100\%$ and the average \lcs\ is used, then 
$L_{gal}({\rm CS54})$ has an upper limit of $31$ \lsun . 
Therefore, the total galactic \lcs\ is likely between $11 - 31$ \lsun\ with
a value most likely near $20$ \lsun .
Assuming that \csfifo\ emission is confined to dense cores within 
molecular clouds (see Helfer \& Blitz 1997),
this estimate of the galactic \lcs\ is probably complete. 
This is consistent with previous estimates of the galactic luminosity
from Paper II and is well below the CS luminosities of nearby starburst galaxies
(see Table 8 in Paper II).

\subsection{Star Formation Rate per Unit Mass}

	The ratio of bolometric luminosity to virial mass is roughly
proportional to the star formation rate per unit mass.  The bolometric
luminosity is calculated from fluxes collected in Table 2 of
Mueller et al. (2002). 
The average \ltom\ ratio is $314$ \lsun $/$\msun, 
ranging from 3 to 2290 \lsun $/$\msun\ for a subsample of $40$ cores
with sufficient flux information to calculate \lbol .  
This average is somewhat higher than those computed for our subsample with
masses from dust emission ($136\pm 100$) and from the sample of 
Beuther et al. (2002) ($120 \pm 90$), once similar assumptions about
dust opacity are made (Mueller et al. 2002b).
We can compare to the values in Paper II after
correcting the Paper II virial mass for density gradients and optically
thick \cs\ linewidths to find \ltom\ $= 440 \pm 100$ \lsun $/$\msun .
The \ltom\ ratio was higher for Paper II due to a bias towards the 
most luminous high-mass star forming regions.  

Sridharan et al. (2002) found that their sample
of sources with low radio continuum emission had a systematically lower
\ltom\ than did a sample of regions with UC\HII\ regions (Hunter et al. 2000).
They interpreted this difference as an evolutionary effect: the sources
without well-developed \HII\ regions were younger and had yet to reach
their full luminosity. Our sample provides a good check of this hypothesis
because it was chosen without regard for the presence of an \HII\ region. 
The majority of the cores in the \ltom\ distribution are associated 
with \HII\ regions
($43\%$) or UC\HII\ regions ($38\%$).  The \ltom\ ratio for cores with 
\HII\ regions is higher (\median\ $= 258$ \lsun $/$\msun )
than for cores with UC\HII\ regions (\median\ $= 166$ \lsun $/$\msun ),
and higher still than for cores without any known radio continuum
(\median\ = $103$ \lsun $/$\msun ).  The distribution of $L/M$ is
plotted in Figure 15i. Thus, our data provide some support for the
interpretation by Sridharan et al. (2002), but the difference is not great,
the overlap of the three samples is substantial, and we have a small 
number of cores without radio continuum emission. 
The \ltom\ for cores with UC\HII\ regions is 1.6 times that
for the sample without, similar to the enhancement of the sample of \HII\
regions studied by Hunter et al. (2000) over that studied by Sridharan
et al. (2002), according to the analysis of those samples by Mueller et al.
(2002). 

All these ratios are much higher than the \ltom\ for molecular clouds, 
as traced by CO generally ($0.4 \lsun /\msun$; Bronfman et al. 2000)
or the enhanced value for molecular clouds with bright \HII\ regions
($4 \lsun /\msun$; Mooney \& Solomon 1988). The dispersion
in this ratio is also less than that for studies using CO, again indicating
that the dense cores are the relevant entities for the study of massive
star formation. This result agrees with studies of HCN toward galaxies
that show a tight, linear relation between far-infrared luminosity and
luminosity of HCN emission (Solomon, Downes, \& Radford 1992; 
Gao \& Solomon 2002).  Those authors argue that the global star formation rate 
per unit mass depends on the fraction of molecular gas in a dense phase. 
We see the same trend in dense cores
in our Galaxy, suggesting that studies of these dense cores may provide
information on conditions in galaxies with intense star formation.

A strong correlation ($r =0.75$) between bolometric luminosity and virial mass
is observed for our sample of cores (Figure 21),
$\log \lbol = (1.70 \pm 0.83) + (1.19 \pm 0.11) \log \mv $.  
The corresponding trend for CO clumps
is shown as a dashed line with slightly flatter slope.
There is no trend in \ltom\ versus \mv\ over two orders of magnitude in
virial mass (Figure 21).  This result is very similar to the 
lack of correlation seen for CO clumps over four order of magnitude in mass
(see Evans 1991), except that the dispersion in \ltom\ for the CS cores in
this survey is a factor of $6$ smaller than for CO clumps.  In the dense
cores within molecular clouds, the star formation rate per unit mass
does not depend on the mass of the core.

Because the luminosity is strongly affected by the
most massive star [$L \propto M_{\star}^{\alpha}$ with $\alpha \sim 3.5$ 
up to $M_{\star} \sim 60$ \msun\ (Scalo 1986)],
the linear relation  between luminosity and
mass and modest dispersion about the relation 
suggests that the mass of the most massive star
is closely related to the mass of the core, with a relation
that approximates $M_{\star}(max) \sim \mv^{1/3.5}$.
Because the mass of the most massive star must be subject to strong
statistical fluctuations, the dispersion in \ltom\ is surprisingly
small; a factor of 2 change in the mass of the most massive star
will cause a change of a factor of 11 in luminosity about the dispersion
that we observe.  Sridharan et al. (2002) come to a similar conclusion
based on their sample of source without UC\HII\ regions.

\subsection{Galactic Trends}

The core size, linewidth, virial mass, surface density, CS abundance,
and luminosity-to-mass ratio
are plotted versus galactic radius in Figure 22.  
The large spike in core sizes near $D_g = 10$ kpc is due to the massive cores 
observed towards the W49 and G32.80+0.20 star forming regions.  
There is little evidence for a trend in core size ($r =-0.01$) or 
linewidth ($r =-0.14$). There may be weak anticorrelations of
surface density ($r =-0.20$), virial mass ($r =-0.26$), 
and luminosity-to-mass ratio ($r =-0.26$) with $D_g$.  
The strongest, but still weak, correlation is between $\log$ CS abundance
and galactocentric radius ($r = -0.32$).
These results mostly agree with previous
CS surveys, which found few trends with galactocentric
distance (e.g., Zinchencko 1995, Zinchencko et al. 1998). 
In particular, Zinchenko et al. (1998) also noted a weak correlation
($r = -0.35$) of $L/M$ with $D_g$.  

Zinchenko et al. (1998) found
that the most significant correlation in their sample was a decrease
of mean density with $D_g$. Their mean densities were obtained 
from the CS column densities, determined from the CS $J=2-1$ line,
assuming an abundance of CS that is constant with $D_g$. 
In contrast, we find no evidence for a decrease in the density determined
from the LVG modeling in Paper II with $D_g$ ($r = -0.03$).
We do see an anticorrelation in CS abundance with
$D_g$ at about the same level of significance as the correlation
Zinchenko et al. found in mean density. A decrease
in abundance could have introduced an artificial decrease in their
mean densities because they assumed a constant abundance.
A decrease in abundance of CS could be caused by many factors,
but a simple explanation would be a Galactic gradient in sulfur abundance,
as has been found by Rudolph et al. (1997).

\section{Conclusions}

We have mapped 63 high-mass star forming cores associated with
water masers in \csfifo .  The source size, aspect ratio,
virial mass, surface density, CS(5--4) luminosity, and \ltom\
ratio were calculated.  A statistical summary of all calculated
quantities is shown in Table 10.  Typically, smaller average
sizes and masses are found compared to results from Paper II
due to the inclusion of weaker sources.

Our main conclusions are as follows:

(1)  A strong correlation is observed between the integrated
intensity of the \csfifo\ line and continuum flux observed at 350 \micron\
(Mueller et al. 2002b) indicating that high-J CS emission
is an excellent tracer of dense gas in high-mass star forming cores.

(2)  The median size is $0.32$ pc.  While a power law density
profile does not have a characteristic size, the median 
FWHM size is comparable
to the core size from dust emission ($R_n$)
determined by Mueller et al. (2002), based on setting the outer radius
at the point where the density drops to $\eten4$ \cmv.

(3)  Most of the core aspect ratios are consistent with spherical
symmetry.  No trend is seen in aspect ratio with $p$, the exponent in
the power law density distribution.

(4)  A weak trend between deconvolved source size and $p$ is observed, as 
expected for power law density profiles.

(5)  There is a very weak correlation between linewidth and size
that is consistent with $\Delta v \propto \rcs ^{0.3}$.  The linewidths of the
cores in this sample are much larger than would be predicted from the 
Caselli \& Myers size-linewidth relation (1995), indicative of high turbulence.

(6)  The median virial mass is $920$ \msun\ after corrections using
\css\ linewidth and $p$.  On average, the virial mass is 2 to 3 times larger
than the mass calculated from 350 \micron\ dust emission toward the same region.

(7)  Source without known radio continuum emission have median CS intensities,
sizes, and masses that are smaller than sources associated with UC\HII , C\HII ,
or \HII\ regions.

(8)  The cumulative mass spectrum is steeper ($\Gamma = -0.9 \pm 0.2$)
than studies of molecular clouds as a whole and clumps within those clouds. 
It is flatter than the Salpeter IMF,
but similar to that of the IMF of OB associations and the distribution of
total masses of stars in OB associations.

(9)  The median pressure of the sample is $1.5 \times 10^8$ K \cmv .  The
high pressure may ameliorate the long standing lifetime problem for  
confinement of Ultra-Compact \HII\ regions.

(10)  The \ltom\ ratio is about two orders of magnitude higher than estimates 
made from tracers of lower density gas (CO) and has a smaller dispersion,
indicating that dense cores traced by submillimeter continuum and high-J CS
emission are the relevant entities for assessing 
the star formation rate per unit mass.  The \ltom\ ratio is $1.6$ times larger
for cores with UC\HII\ regions compared to cores without UC\HII\ regions.

(11) A strong correlation is observed between luminosity and virial mass.
This result combined with the low dispersion in \ltom\ indicates that
the mass of the most massive star is likely related to the mass of the core.

(12)  No trends in size, mass, or 
\ltom\ with galactocentric radius are apparent.  
A weak decrease in CS abundance with galactocentric distance
is observed.

\section*{Acknowledgments}

We are grateful to the staff of the CSO for excellent assistance,
to Richard Chamberlin for assistance with beam information,
and to Jeong-Eun Lee for help compiling the CSO efficiencies and pointing
measurements.
We thank C. McKee, J. Tan, and P. Solomon for helpful discussions.
This research has made use of the SIMBAD database, 
operated at CDS, Strasbourg, France.  
This work was supported by NSF Grant AST-9988230 and the State of Texas.





\begin{figure}
\figurenum{1}
\epsscale{0.8}
\figcaption{
Histogram of distances and the position of the 63 mapped cores in the 
galactic plane.  The histogram is binned at 2 kpc.  The median (dotted line)
and mean (dashed line) are shown. In the galactic coordinates plot,
the Sun is at the center.  Since the observations were
performed at the CSO in the northern hemisphere, almost all of the cores are in the
first and second quadrant.  The circles represent distances of 5 kpc and 10 kpc from
the Sun.}
\end{figure}

\begin{figure}
\figurenum{2}
\epsscale{0.8}
\figcaption{
Contour maps of \csfifo\ integrated intensity with the FWHM beamsize shown in the 
lower left corner. The contour levels are indicated at the bottom of each
panel.  For instance, $5\% ,10\% (5\sigma )$ means the first contour is 5\% the
peak intensity, the next contour is 10\% the peak intensity, and the
contour interval is 10\% or 5$\sigma$. 
The plus sign marks the location of the nearest \HII\ region to the
water maser position. The water maser is at (0,0).
}
\end{figure}

\begin{figure}
\figurenum{3}
\epsscale{0.8}
\figcaption{
Contour maps of \csfifo\ integrated intensity with the FWHM beamsize shown in the 
lower left corner. The contour levels are indicated at the bottom of each
panel.  For instance, $5\% ,10\% (5\sigma )$ means the first contour is 5\% the
peak intensity, the next contour is 10\% the peak intensity, and the
contour interval is 10\% or 5$\sigma$. 
The plus sign marks the location of the nearest \HII\ region to the
water maser position. The water maser is at (0,0).
}
\end{figure}

\begin{figure}
\figurenum{4}
\epsscale{0.8}
\figcaption{
Contour maps of \csfifo\ integrated intensity with the FWHM beamsize shown in the 
lower left corner. The contour levels are indicated at the bottom of each
panel.  For instance, $5\% ,10\% (5\sigma )$ means the first contour is 5\% the
peak intensity, the next contour is 10\% the peak intensity, and the
contour interval is 10\% or 5$\sigma$. 
The plus sign marks the location of the nearest \HII\ region to the
water maser position. The water maser is at (0,0).
}
\end{figure}

\begin{figure}
\figurenum{5}
\epsscale{0.8}
\figcaption{
Contour maps of \csfifo\ integrated intensity with the FWHM beamsize shown in the 
lower left corner. The contour levels are indicated at the bottom of each
panel.  For instance, $5\% ,10\% (5\sigma )$ means the first contour is 5\% the
peak intensity, the next contour is 10\% the peak intensity, and the
contour interval is 10\% or 5$\sigma$. 
The plus sign marks the location of the nearest \HII\ region to the
water maser position. The water maser is at (0,0).
}
\end{figure}

\begin{figure}
\figurenum{6}
\epsscale{0.8}
\figcaption{
Contour maps of \csfifo\ integrated intensity with the FWHM beamsize shown in the 
lower left corner. The contour levels are indicated at the bottom of each
panel.  For instance, $5\% ,10\% (5\sigma )$ means the first contour is 5\% the
peak intensity, the next contour is 10\% the peak intensity, and the
contour interval is 10\% or 5$\sigma$. 
The plus sign marks the location of the nearest \HII\ region to the
water maser position. The water maser is at (0,0).
}
\end{figure}

\begin{figure}
\figurenum{7}
\epsscale{0.8}
\figcaption{
Contour maps of \csfifo\ integrated intensity with the FWHM beamsize shown in the 
lower left corner. The contour levels are indicated at the bottom of each
panel.  For instance, $5\% ,10\% (5\sigma )$ means the first contour is 5\% the
peak intensity, the next contour is 10\% the peak intensity, and the
contour interval is 10\% or 5$\sigma$. 
The plus sign marks the location of the nearest \HII\ region to the
water maser position. The water maser is at (0,0).
}
\end{figure}

\begin{figure}
\figurenum{8}
\epsscale{0.8}
\figcaption{
Contour maps of \csfifo\ integrated intensity with the FWHM beamsize shown in the 
lower left corner. The contour levels are indicated at the bottom of each
panel.  For instance, $5\% ,10\% (5\sigma )$ means the first contour is 5\% the
peak intensity, the next contour is 10\% the peak intensity, and the
contour interval is 10\% or 5$\sigma$. 
The plus sign marks the location of the nearest \HII\ region to the
water maser position. The water maser is at (0,0).
}
\end{figure}

\begin{figure}
\figurenum{9}
\epsscale{0.8}
\figcaption{
Contour maps of \csfifo\ integrated intensity with the FWHM beamsize shown in the 
lower left corner. The contour levels are indicated at the bottom of each
panel.  For instance, $5\% ,10\% (5\sigma )$ means the first contour is 5\% the
peak intensity, the next contour is 10\% the peak intensity, and the
contour interval is 10\% or 5$\sigma$. 
The plus sign marks the location of the nearest \HII\ region to the
water maser position. The water maser is at (0,0).
}
\end{figure}

\begin{figure}
\figurenum{10}
\epsscale{0.8}
\figcaption{
Contour maps of \csfifo\ integrated intensity with the FWHM beamsize shown in the 
lower left corner. The contour levels are indicated at the bottom of each
panel.  For instance, $5\% ,10\% (5\sigma )$ means the first contour is 5\% the
peak intensity, the next contour is 10\% the peak intensity, and the
contour interval is 10\% or 5$\sigma$. 
The plus sign marks the location of the nearest \HII\ region to the
water maser position. The water maser is at (0,0).
}
\end{figure}

\begin{figure}
\figurenum{11}
\epsscale{0.8}
\figcaption{
Contour maps of \csfifo\ integrated intensity with the FWHM beamsize shown in the 
lower left corner. The contour levels are indicated at the bottom of each
panel.  For instance, $5\% ,10\% (5\sigma )$ means the first contour is 5\% the
peak intensity, the next contour is 10\% the peak intensity, and the
contour interval is 10\% or 5$\sigma$. 
The plus sign marks the location of the nearest \HII\ region to the
water maser position. The water maser is at (0,0) except in the W75(OH)/DR21S
map where a second water maser is marked by a triangle near the peak of DR21S.
}
\end{figure}

\begin{figure}
\figurenum{12}
\epsscale{0.8}
\figcaption{
Contour maps of \csfifo\ integrated intensity with the FWHM beamsize shown in the 
lower left corner. The contour levels are indicated at the bottom of each
panel.  For instance, $5\% ,10\% (5\sigma )$ means the first contour is 5\% the
peak intensity, the next contour is 10\% the peak intensity, and the
contour interval is 10\% or 5$\sigma$. 
The plus sign marks the location of the nearest \HII\ region to the
water maser position. The water maser is at (0,0).
}
\end{figure}

\begin{figure}
\figurenum{13}
\epsscale{0.8}
\figcaption{
The logarithm of the peak CS 5-4 integrated intensity vs. the logarithm
of the 350 \micron\ flux density in a 30\as\ aperture (Mueller et al. 2002).  
The solid line indicates the best fit relation:
$\log I(T_R^*)$ $= (-0.60 \pm 0.01) + (0.92 \pm 0.05)\log S_{350 \micron}$.
}
\end{figure}

\begin{figure}
\figurenum{14}
\epsscale{0.8}
\figcaption{
Plot of core size vs. distance (upper panel) and integrated intensity vs. \rcs\
(lower panel). The dashed
line in the upper panel shows the size of a core with a deconvolved size equal to
the beamsize while the horizontal dashed line markes the source size with
a deconvolved source size equal to the beam for the median distance of the sample.
No correlation is observed between 
integrated intensity and \rcs .}
\end{figure}

\begin{figure}
\figurenum{15}
\epsscale{0.9}
\figcaption{
Histograms of \rcs , \ar , major axis position angle, M$_{vir}$, \surfden , 
\cs\ abundance, \csfifo\ luminosity, and \ltom .  Distributions of sources without 
a known radio continuum detection are plotted as dashed-line histograms 
(panels a, d, f, g, h, and i).  The total
distributions (including all sources) are plotted as solid-line histograms.
The mean (dashed vertical line) and 
median (dotted vertical line) of the total distributions are plotted.
}
\end{figure}

\begin{figure}
\figurenum{16}
\epsscale{0.8}
\figcaption{
The aspect ratio (lower panel) and the deconvolved source
size ($\theta _{dec}/\theta _{mb}$; upper panel) are compared to the best fit
power law density index determined by Mueller et al. 2002.
No correlation between aspect ratio
and $p$ is observed while a weak correlation is observed between $p$ and
$\theta _{dec}/\theta _{mb}$.  The typical errorbars for $(a/b)_{obs}$ and $p$ 
are shown in the bottom panel.}
\end{figure}

\begin{figure}
\figurenum{17}
\epsscale{0.8}
\figcaption{
The upper panel
shows the CS linewidth compared with C$^{34}$S linewidth.  For the subsample of
cores mapped in both CS and C$^{34}$S, the CS linewidth is broader on average.
The lower panel shows the C$^{34}$S and $^{13}$CS linewidths. }
\end{figure}

\begin{figure}
\figurenum{18}
\epsscale{0.8}
\figcaption{
The linewidth-size relationship using C$^{34}$S linewidths.  
The FWHM size, \rcs , is shown in the top panel and while the
size at an intensity of 10 K km/s, $R_{10}$, is shown in
the bottom panel.  The extrapolated 
linewidth-size relationships for low and high mass regions are labeled and the least
squares fit and robust estimation for our sample are shown.
}
\end{figure}

\begin{figure}
\figurenum{19}
\epsscale{0.8}
\figcaption{
The logarithm of $I(T_R^*)$ and $\log \mv$ are compared in the upper panel and the
virial mass and dust determined mass are compared in the lower panel.  
More massive cores are typically brighter in CS intensity: $\log I(T_R^*) 
= (-0.76 \pm 0.11) + (0.81 \pm 0.04)\log \mv $. 
The virial mass and mass derived from dust continuum emission correlate well, but
$\mv > M_{dust}$.  The solid line in the top panel is the least squares fit while 
the solid line in the bottom panel indicates \mv\ $=$ $M_{dust}$.
}
\end{figure}

\begin{figure}
\figurenum{20}
\epsscale{0.8}
\figcaption{
The cumulative mass spectrum determined from the CS core virial mass.  Least
squares and robust estimation fits are shown as well as the Salpeter
IMF and CO clump mass slope.
}
\end{figure}

\begin{figure}
\figurenum{21}
\epsscale{0.8}
\figcaption{
The top panel plots \lbol\ vs. \mvir\ and the 
bottom panel plots 
$L_{bol}/M_{vir}$ vs. the \mvir.  Source with \HII\ regions
are plotted as open triangles,  sources with UC\HII\ regions
are plotted as open circles, and source without a known UC\HII\
region are plotted as filled squares.
The dotted line in the top panel
is the relationship derived for CO clumps while the solid line is a
least squares fit.
$L_{bol}/M_{vir}$ is proportional
to the star formation rate per unit mass.   The range of $L_{bol}/M_{vir}$
for CO clumps is shown as a double arrow at the left of the bottom
panel.  The dispersion observed towards CS cores is roughly 6 times smaller 
than the equivalent relationship for CO clumps (Evans 1991).
}
\end{figure}

\begin{figure}
\figurenum{22}
\epsscale{0.8}
\figcaption{
Plot of \rcs , \dvs , M$_{vir}$, \surfden , $X(CS)$, 
and L$_{bol}$/M$_{vir}$ versus galactocentric
distance.  
Source with \HII\ regions
are plotted as open triangles,  sources with UC\HII\ regions
are plotted as open circles, and source without a known UC\HII\
region are plotted as filled squares.
Only the CS abundance shows a weak correlation with galactocentric
distance.}
\end{figure}


\begin{deluxetable}{lllccccccc}
\tablecolumns{10}
\tiny
\tablecaption{Observed Sources \label{tab1}}
\tablewidth{0pt} 
\tablehead{
\colhead{Source}                  &
\colhead{$\alpha$ (1950.0)}       &
\colhead{$\delta$ (1950.0)}       &
\colhead{Date \cs }               &
\colhead{Dist.}                   &
\colhead{Dist.}                   &
\colhead{D$_g$} 		  &
\colhead{\HII\ ?}		  &
\colhead{\HII\ } 		  &
\colhead{Size } 		  \\
\colhead{}                                &
\colhead{($^h$~~$^m$~~$^s$~)~}            &  
\colhead{($\degree$ ~\arcmin\ ~\arcsec)}  &  
\colhead{Mapped}                          &
\colhead{(kpc)}                           &
\colhead{Ref.}                            &
\colhead{(kpc)}	     			  & 
\colhead{}				  &
\colhead{R$_{2 cm}$ (pc)}			  & 
\colhead{Ref.}	    
}
\startdata 
G121.30+0.66    & 00 33 53.3 & $+$63 12 32 & 12/97 & 1.2  &  1 & 9.2 	& ...     & ...  & \\
G123.07-6.31    & 00 49 29.2 & $+$56 17 36 & 12/97 & 2.2  &  2 & 9.9 	& \HII\	  & ... \\
W3(OH)		& 02 23 17.3 & $+$61 38 58 & 12/96 & 2.4  &  3 & 10.3 	& UC\HII\ & 0.02 & 18  \\
G135.28+2.80	& 02 39 31.0 & $+$62 44 16 & 12/97 & 7.4  &  1 & 14.7 	& ...	  & ... \\
S231          	& 05 35 51.3 & $+$35 44 16 & 12/96 & 2.3  &  2 & 10.8 	& ...	  & ... \\
S235		& 05 37 31.8 & $+$35 40 18 & 12/96 & 1.6  &  2 & 10.1 	& \HII\	  & ... \\
S241		& 06 00 40.9 & $+$30 14 54 & 12/97 & 4.7  &  2 & 13.2 	& ...	  & ... \\
S252A		& 06 05 36.5 & $+$20 39 34 & 12/97 & 1.5  &  2 & 10.0 	& \HII\	  & ... \\
S255		& 06 09 58.3 & $+$18 00 12 & 12/96 & 1.3  &  4 & 9.8 	& UC\HII\ & 0.01 & 19 \\
RCW142          & 17 47 04.5 & $-$28 53 42 & 04/97 & 2.0  &  5 & 6.5 	& UC\HII\ & ...  & \\
W28A2(1)        & 17 57 26.8 & $-$24 03 54 & 09/96 & 2.6  &  5,6 & 5.9  & UC\HII\ & 0.05 & 14 \\
M8E             & 18 01 49.1 & $-$24 26 57 & 04/97 & 1.8  &  2 & 6.7  	& UC\HII\ & ... \\
G9.62+0.10	& 18 03 16.0 & $-$20 32 01 & 09/96 & 7.0  &  7 & 3.0 	& UC\HII\ & 0.02 & 19 \\
G8.67-0.36      & 18 03 18.6 & $-$21 37 59 & 04/97 & 8.5  &  8 & 4.1 	& UC\HII\ & 0.03 & 14 \\
W31		& 18 05 40.4 & $-$19 52 21 & 09/96 & 12.0 &  4 & 4.0 	& UC\HII\ & 0.05 & 20 \\
G10.6-0.4       & 18 07 30.7 & $-$19 56 28 & 09/96 & 6.5  &  9 & 2.4 	& C\HII\  & 0.06 & 14 \\
G12.42+0.50	& 18 07 56.4 & $-$17 56 37 & 04/97 & 2.1  &  10 & 6.5 	& UC\HII\ & 0.01 & 21 \\
G12.89+0.49     & 18 08 56.3 & $-$17 53 09 & 04/97 & 3.5  &  8 & 5.1 	& ...	  & ... \\
G12.2-0.1       & 18 09 43.7 & $-$18 25 09 & 09/96 & 16.3 &  11 & 5.7 	& C\HII\  & 0.27 & 14 \\
W33cont         & 18 11 18.3 & $-$17 56 21 & 10/96 & 4.1  &  9 & 4.6 	& UC\HII\ & ... \\
G13.87+0.28     & 18 11 41.5 & $-$16 16 34 & 07/98 & 4.4  &  12 & 4.4 	& \HII\   & 0.41 & 20 \\
W33A		& 18 11 44.0 & $-$17 53 09 & 04/97 & 4.5  &  5 & 4.2 	& ...	  & ... \\
G14.33-0.64	& 18 16 00.8 & $-$16 49 06 & 04/97 & 2.6  &  8 & 6.0 	& UC\HII\ & ... \\
G19.61-0.23     & 18 24 50.1 & $-$11 58 22 & 09/96 & 4.0  &  4 & 4.9 	& C\HII\  & 0.12 & 14	\\
G20.08-0.13	& 18 25 22.6 & $-$11 30 45 & 07/98 & 3.4  &  7 & 5.4 	& UC\HII\ & 0.05 & 14	\\
G23.95+0.16     & 18 31 40.8 & $-$16 16 34 & 07/98 & 5.8  &  9 & 4.0 	& \HII\   & 0.32 & 14 \\
G24.49-0.04     & 18 33 22.8 & $-$07 33 54 & 04/97 & 3.5  &  1 & 5.5 	& ...     & ...	  & ... \\
W42		& 18 33 30.3 & $-$07 14 42 & 04/97 & 9.1  &  13 & 3.8 	& UC\HII\ & ... \\
G28.86+0.07	& 18 41 07.9 & $-$03 38 41 & 07/98 & 8.5  &  5 & 4.2 	& ...  	  & ...   & ... \\
W43S		& 18 43 26.7 & $-$02 42 40 & 07/98 & 8.5  &  4,14 & 4.4 & C\HII\  & 0.28 & 14 \\
G31.41+0.31     & 18 44 59.5 & $-$01 16 07 & 04/97 & 7.9  &  12 & 4.5 	& UC\HII\ & 0.05 & 14 \\
\enddata
\tablerefs{\footnotesize 1. R$_N$ Paper I 1992, 2. Blitz 1982, 3. Harris 1976, 4.Genzel 1977,
5. Braz 1983, 6. Chini 1986, 7. Hofner 1996, 8. Val'tts 2000, 9. Solomon 1987, 
10. Zinchecnko 1994, 11. Hunter 2000, 12. Churchwell 1990, 13. Downes 1980,
14. Wood 1989, 15. Brand 1993, 16. Wink 1982, 17. Zhou 1996, 18. Wilner 1995, 
19. Kurtz 1994, 20. Hatchell 2000, 21. Jaffe 1984}
\end{deluxetable}

\begin{deluxetable}{lllccccccc}
\tablecolumns{10}
\tiny
\tablecaption{Observed Sources Cont. \label{tab2}}
\tablewidth{0pt} 
\tablehead{
\colhead{Source}                  &
\colhead{$\alpha$ (1950.0)}       &
\colhead{$\delta$ (1950.0)}       &
\colhead{Date \cs }               &
\colhead{Dist.}                   &
\colhead{Dist.}                   &
\colhead{D$_g$} 		  &
\colhead{\HII\ ?}		  &
\colhead{\HII\ }		  &
\colhead{Size} 		  \\
\colhead{}                                &
\colhead{($^h$~~$^m$~~$^s$~)~}            &  
\colhead{($\degree$ ~\arcmin\ ~\arcsec)}  &  
\colhead{Mapped}                          &
\colhead{(kpc)}                           &
\colhead{Ref.}                            &
\colhead{(kpc)}	     			  &
\colhead{}				  & 
\colhead{R$_{2 cm}$ (pc)}	&
\colhead{Ref.}	    
}
\startdata 
W43Main3	& 18 45 11.2 & $-$01 57 57 & 07/98 & 6.8  &  4 & 4.4	& ...     & ...	& ...\\
G31.44-0.26     & 18 46 57.5 & $-$01 32 33 & 04/97 & 10.7 &  9 & 5.6 	& UC\HII\ & 0.04 & 19	\\
G32.05+0.06	& 18 47 02.0 & $-$00 49 19 & 07/98 & 8.5  &  9 & 4.7 	& ...	  & ... \\
G32.80+0.20A/B  & 18 47 57.3 & $-$00 05 28 & 07/98 & 15.6 & 13 & 9.6 	& C\HII\  & 0.09 & 19	\\
W44		& 18 50 46.1 & $+$01 11 11 & 07/98 & 3.7  &  9 & 5.8 	& C\HII\  & 0.06 & 14	\\
S76E		& 18 53 45.6 & $+$07 49 16 & 07/98 & 2.1  &  1 & 7.0 	& \HII\	  & ... \\
G35.58-0.03     & 18 53 51.4 & $+$02 16 29 & 10/96 & 3.5  &  13 & 6.0 	& UC\HII\ & 0.02 & 19 \\
G35.20-0.74     & 18 55 40.8 & $+$01 36 30 & 07/98 & 3.3  &  9 & 6.1 	& \HII\	  & ... \\
W49N		& 19 07 49.8 & $+$09 01 17 & 10/96 & 14.0 &  4 & 9.7 	& UC\HII\ & 0.01 & 22 \\
W49S		& 19 07 58.2 & $+$09 00 03 & 07/99 & 14.0 &  4 & 9.7 	& UC\HII\ & ... \\
OH43.80-0.13	& 19 09 31.2 & $+$09 30 51 & 07/98 & 2.7  & 13 & 6.8 	& UC\HII\ & 0.01 & 19 \\
G45.07+0.13     & 19 11 00.3 & $+$10 45 42 & 09/96 & 9.7  &  13 & 7.1 	& UC\HII\ & 0.04 & 14 \\
G48.61+0.02     & 19 18 13.1 & $+$13 49 44 & 07/98 & 11.8 &  1  & 8.9 	& C\HII\  & 0.07 & 19 \\
W51W		& 19 20 53.3 & $+$14 20 47 & 07/99 & 7.0  &  17 & 6.6 	& \HII\	  & ... \\
W51M		& 19 21 26.2 & $+$14 24 36 & 10/96 & 7.0  &  17 & 6.6 	& C\HII\  & 0.21 & 23 \\
G59.78+0.06	& 19 41 04.2 & $+$23 36 42 & 07/98 & 2.2  &  1  & 7.6 	& UC\HII\ & ...	  & ... \\
S87		& 19 44 14.0 & $+$24 28 10 & 09/96 & 1.9  &  15 & 7.6 	& UC\HII\ & 0.01 & 19 \\
S88B		& 19 44 42.0 & $+$25 05 30 & 07/96 & 2.1  &  2  & 7.7 	& UC\HII\ & 0.01 & 14 \\
K3-50		& 19 59 50.1 & $+$33 24 17 & 06/97 & 9.0  &  4  & 10.1 	& C\HII\  & 0.18 & 19 \\
ON1		& 20 08 09.9 & $+$31 22 42 & 07/98 & 6.0  &  1  & 8.5 	& UC\HII\ & 0.02 & 19 \\
ON2S		& 20 19 48.9 & $+$37 15 52 & 07/98 & 5.5  &  1  & 8.9 	& \HII\	  & ... \\
ON2N		& 20 19 51.8 & $+$37 17 01 & 07/98 & 5.5  &  1  & 8.9 	& C\HII\  & 0.07 & 14 \\
S106            & 20 25 32.8 & $+$37 12 54 & 07/98 & 4.1  & 16  & 8.5 	& UC\HII\ & 0.01 & 19 \\
W75N		& 20 36 50.5 & $+$42 27 01 & 07/98 & 3.0  &  4  & 8.6 	& UC\HII\ & ... \\
DR21S		& 20 37 13.8 & $+$42 08 52 & 07/99 & 3.0  &  4  & 8.6 	& UC\HII\ & 0.04 & 19 \\
W75(OH)		& 20 37 14.1 & $+$42 12 12 & 07/99 & 3.0  &  4  & 8.6 	& ...	  & ... \\
G97.53+3.19	& 21 30 37.0 & $+$55 40 36 & 07/98 & 7.9  &  1  & 12.3 	& \HII\	  & ... \\
BFS11-B		& 21 41 57.6 & $+$65 53 17 & 12/97 & 2.0  &  5  & 9.2 	& ...     & ... & ... \\
CepA		& 22 54 19.2 & $+$61 45 44 & 07/99 & 0.73 & 17  & 8.8 	& UC\HII\ & ... & ... \\
NGC7538		& 23 11 36.1 & $+$61 10 30 & 12/97 & 2.8  &  2  & 9.9 	& UC\HII\ & $<$0.01 & 24 \\
S157		& 23 13 53.1 & $+$59 45 18 & 12/97 & 2.5  &  2  & 9.7 	& C\HII\  & 0.10 & 19 \\ 
\enddata
\tablerefs{\footnotesize 1. R$_N$ Paper I 1992, 2. Blitz 1982, 3. Harris 1976, 4.Genzel 1977,
5. Braz 1983, 6. Chini 1986, 7. Hofner 1996, 8. Val'tts 2000, 9. Solomon 1987, 
10. Zinchecnko 1994, 11. Hunter 2000, 12. Churchwell 1990, 13. Downes 1980,
14. Wood 1989, 15. Brand 1993, 16. Wink 1982, 17. Zhou 1996, 18. Wilner 1995, 
19. Kurtz 1994, 20. Hatchell 2000,
21. Jaffe 1984, 22. Dreher 1984, 23. Scott 1978, 24. Turner 1984}
\end{deluxetable}

\begin{deluxetable}{lrcccc}
\tablecolumns{6}
\footnotesize
\tablecaption{CSO Observations 1996-2002\label{tab3}}
\tablewidth{0pt} 
\tablehead{
\colhead{UT Date}                 &
\colhead{Transition}		  &
\colhead{$\nu$}			  &
\colhead{$\theta_{mb}$}           &
\colhead{$\eta_{mb}$}             &
\colhead{Pointing}                \\
\colhead{}                        &
\colhead{}			  &
\colhead{(GHz)}			  &
\colhead{(\as )}                  &  
\colhead{}                        &  
\colhead{($\sigma _{AZ}$\as\ ,$\sigma _{ZA}$\as\ )}  
}
\startdata 
09/96	& \csfifo\ & 244.9355680	& 24.5 & 0.56 & (3.8,2.7)\\
10/96	& \csfifo\ & 244.9355680	& 24.5 & 0.56 & (3.8,2.7)\\
12/96	& \csfifo\ & 244.9355680	& 24.5 & 0.54 & (1.9,3.2) \\
04/97	& \csfifo\ & 244.9355680	& 24.5 & 0.56 & (6.6,3.2)\\
06/97	& \csfifo\ & 244.9355680	& 24.5 & 0.58 & (2.0,5.2) \\
12/97	& \csfifo\ & 244.9355680	& 24.5 & 0.55 & (4.4,5.2) \\
07/98	& \csfifo\ & 244.9355680	& 24.5 & 0.57 & (6.7,2.7) \\
12/98	& \cssfifo\ & 241.0161940	& 31.0 & 0.66 & (2.4,3.4) \\
	& \ccsfifo\ & 231.2209960	& 32.5 & 0.66 &  \\
07/99	& \csfifo\ & 244.9355680	& 30.5 & 0.64 & (5.6,5.4) \\
07/01	& \cssfifo\ & 241.0161940	& 31.0 & 0.73 & (3.0,3.4) \\
	& \ccsfifo\ & 231.2209960	& 32.5 & 0.73 &  \\
01/02	& \cssfifo\ & 241.0161940	& 31.0 & 0.54 & (12.7,3.8) \\ 
06/02	& \cssfifo\ & 241.0161940  	& 31.0 & 0.61 & (4.2,4.8) \\ 
	& \ccsfifo\ & 231.2209960	& 32.5 & 0.59 &  \\
\enddata
\end{deluxetable}

\begin{deluxetable}{lcccccc}
\tablecolumns{7}
\footnotesize
\tablecaption{Observed Line Parameters\label{tab4}}
\tablewidth{0pt} 
\tablehead{
\colhead{Source}                  &
\colhead{I(\tr , \cs )\tablenotemark{a}} &
\colhead{$\Delta v$(CS)}    &
\colhead{I(\tr , \css )\tablenotemark{a}} &
\colhead{$\Delta v$(C$^{34}$S)}    &
\colhead{I(\tr , \ccs )\tablenotemark{a}} &
\colhead{$\Delta v$($^{13}$CS)}    \\
\colhead{}                        &
\colhead{(K \kms)}                &  
\colhead{(\kms)}                  &
\colhead{(K \kms)}                &  
\colhead{(\kms)}  		  &
\colhead{(K \kms)}                &  
\colhead{(\kms)}  			  
}
\startdata 
G121.30+0.66    & 22.2 (2.3)  & 3.46 (0.13) & 1.5 (0.2) & 4.23 (0.39) 	& ...	& ... \\
G123.07-6.31    & 25.8 (2.7)  & 4.49 (0.13) & 1.1 (0.2)	& 4.89 (0.45)	& ...	& ... \\
W3(OH)		& 72.4 (7.3)  & 5.92 (0.13) & 6.3 (0.7) & 5.80 (0.18)	& 2.5 (0.3) & 4.92 (0.25) \\
G135.28+2.80	& 6.5 (0.8)   & 3.46 (0.13) & ...	& ...		& ...	& ... \\
S231          	& 27.4 (2.8)  & 3.89 (0.13) & 1.6 (0.2) & 2.48 (0.19)	& ...	& ... \\
S235		& 32.0 (3.3)  & 2.68 (0.12) & 2.9 (0.4) & 2.09 (0.18)	& 0.5 (0.2) & 2.32 (0.30) \\
S241		& 7.5 (0.9)   & 2.63 (0.14) & ...	& ...		& ...	& ... \\
S252A		& 17.1 (1.8)  & 3.11 (0.12) & ...	& ...		& ...	& ... \\
S255		& 47.8 (4.8)  & 3.12 (0.12) & ...	& ...		& ...	& ... \\
RCW142          & 116 (12)    & 6.00 (0.13) & 26.2 (2.7)& 5.60 (0.17)	& 12.8 (1.3) & 5.52 (0.14)  \\
W28A2(1)        & 204 (20)    & 6.85 (0.15) & 23.2 (2.3)& 5.91 (0.15)	& 8.9 (0.9)  & 5.28 (0.16) \\
M8E             & 32.5 (3.3)  & 3.12 (0.12) & 4.7 (0.5) & 2.23 (0.15)	& ...	& ...  \\
G9.62+0.10	& 55.4 (5.6)  & 7.26 (0.19) & 16.0 (1.6)& 7.33 (0.38)	& ...	& ...  \\
G8.67-0.36      & 47.0 (4.8)  & 5.43 (0.15) & 5.6 (0.6) & 5.08 (0.27)	& ...	& ...  \\
W31		& 55.9 (5.7)  & 11.11 (0.26)& 7.3 (0.8) & 8.56 (0.31)	& ...	& ...  \\
G10.6-0.4       & 182 (18)    & 7.04 (0.13)& 29.8 (3.0)& 6.72 (0.14)	& 17.1 (1.7) & 6.43 (0.13)  \\
G12.42+0.50	& 24.8 (3.5)  & 3.13 (0.13) & ...	&  ...		& ... 	& ... \\
G12.89+0.49     & 30.0 (3.1)  & 5.09 (0.13) & 5.4 (0.5) & 3.78 (0.14)	& ...	& ... \\
G12.2-0.1       & 35.0 (3.8)  & 8.01 (0.22) & 4.2 (0.4) & 7.06 (0.22)	& ...	& ... \\
W33cont         & 122 (13)    & 6.49 (0.14)& 21.1 (2.1)& 5.13 (0.13)	& 10.8 (1.0) & 4.72 (0.14) \\
G13.87+0.28     & 17.5 (1.9)  & 4.15 (0.18) & 2.7 (0.3) & 2.50 (0.21) 	& ...	& ...  \\
W33A		& 32.0 (3.3)  & 4.96 (0.18) & 2.3 (0.3) & 3.22 (0.27)	& ...	& ... \\
G14.33-0.64	& 53.6 (5.4)  & 4.97 (0.14) & 4.5 (0.5) & 2.74 (0.14)	& ...	& ... \\
G19.61-0.23     & 53.4 (5.8)  & 8.97 (0.23) & 3.2 (0.4) & 6.50 (0.30)	& ...	& ... \\
G20.08-0.13	& 26.7 (2.7)  & 8.20 (0.16) & 4.5 (0.5) & 8.39 (0.46)	& ...	& ... \\
G23.95+0.16     & 18.1 (1.9)  & 3.01 (0.13) & 3.2 (0.4) & 2.39 (0.25)	& ...	& ... \\
G24.49-0.04     & 17.3 (1.8)  & 4.43 (0.18) & ...	& ...		& ...	& ... \\
W42		& 35.7 (3.7)  & 8.42 (0.19) & 8.7 (0.9) & 5.44 (0.13)	& ...	& ... \\
G28.86+0.07	& 16.3 (1.7)  & 5.34 (0.15) & 2.5 (0.3) & 3.17 (0.19)	& ...	& ... \\
W43S		& 52.6 (7.7)  & 5.01 (0.13) & 8.0 (0.8) & 3.97 (0.13)	& ...	& ... \\
G31.41+0.31     & 44.3 (4.5)  & 5.89 (0.24) & 7.7 (0.8) & 5.86 (0.20)	& ...	& ... \\
\enddata
\tablenotetext{a}{Peak position.}
\end{deluxetable}

\begin{deluxetable}{lcccccc}
\tablecolumns{7}
\footnotesize
\tablecaption{Observed Line Parameters Cont.\label{tab5}}
\tablewidth{0pt} 
\tablehead{
\colhead{Source}                  &
\colhead{I(\tr , \cs )\tablenotemark{a}} &
\colhead{$\Delta v$(CS)}    &
\colhead{I(\tr , \css )\tablenotemark{a}} &
\colhead{$\Delta v$(C$^{34}$S)}    &
\colhead{I(\tr , \ccs )\tablenotemark{a}} &
\colhead{$\Delta v$($^{13}$CS)}    \\
\colhead{}                        &
\colhead{(K \kms)}                &  
\colhead{(\kms)}                  &
\colhead{(K \kms)}                &  
\colhead{(\kms)}  		  &
\colhead{(K \kms)}                &  
\colhead{(\kms)}  			  
}
\startdata 
W43Main3	& 37.5 (4.1)   & 9.68 (0.12) & 7.4 (0.8) &  6.82 (0.23)	& ...	& ... \\
G31.44-0.26     & 22.7 (2.3)   & 5.22 (0.14) & 2.0 (0.2) &  3.80 (0.38)	& ...	& ... \\
G32.05+0.06	& 14.6 (1.5)   & 8.04 (0.18) & 1.9 (0.2) &  4.54 (0.59)	& ...	& ... \\
G32.80+0.20A/B  & 28.1 (2.9)   & 8.04 (0.14) & 1.7 (0.3) &  5.16 (0.74)	& ...	& ... \\
W44		& 107 (11)     & 5.92 (0.12) & 21.8 (2.2)&  5.04 (0.51)	& 7.6 (0.8) & 4.72 (0.14) \\
S76E		& 56.3 (5.7)   & 3.70 (0.12) & ...	 &  ...		& ...	& ...  \\
G35.58-0.03     & 22.7 (2.5)   & 5.01 (0.18) & 2.2 (0.4) &  6.56 (0.52)	& ...	& ... \\
G35.20-0.74     & 31.8 (3.3)   & 6.49 (0.13) & 1.9 (0.2) &  8.45 (0.71)	& ...	& ... \\
W49N 4 \kms	& 103 (11)     & 9.79 (1.38) & 2.2 (0.2) &  9.80 (0.94)	& 1.4 (0.2) & 9.14 (1.77) \\
W49N 12 \kms 	& 63.4 (6.6)   & 9.54 (1.38) & 3.0 (0.7) &  5.60 (0.94)	& 3.8 (0.4) & 7.38 (1.77) \\
W49S		& 27.3 (3.0)   & 8.32 (0.15) & 1.9 (0.2) &  7.56 (0.44)	& ...	& ... \\
OH43.80-0.13	& 28.8 (2.9)   & 7.55 (0.23) & 1.0 (0.1) &  4.12 (0.30)	& ...	& ... \\
G45.07+0.13     & 42.3 (4.3)   & 6.08 (0.16) & 7.8 (0.8) &  6.10 (0.21)	& ...	& ... \\
G48.61+0.02     & 15.1 (1.6)   & 5.00 (0.17) & 0.4 (0.1) &  2.34 (0.24)	& ...	& ... \\
W51W		& 29.1 (3.3)   & 3.82 (0.14) & 4.2 (2.0) &  3.39 (0.28)	& ...	& ... \\
W51M		& 230 (23)     & 10.95 (0.13)& 26.7 (2.7)&  8.96 (0.18)	& 19.2 (1.9) & 8.03 (0.19) \\
G59.78+0.06	& 17.4 (1.9)   & 3.20 (0.15) & 0.6 (0.1) &  1.11 (0.19)	& ...	& ... \\
S87		& 28.6 (3.0)   & 2.49 (0.16) & ...	 &  ...		& ...	& ... \\
S88B		& 21.1 (2.2)   & 3.06 (0.13) & 1.3 (0.1) &  2.35 (0.18)	& ...	& ... \\
K3-50		& 25.2 (2.7)   & 8.07 (0.15) & 1.4 (0.2) &  7.61 (0.56)	& ...	& ... \\
ON1		& 20.2 (2.1)   & 4.68 (0.13) & 2.0 (0.3) &  4.51 (0.24)	& ...	& ...  \\
ON2S		& 42.3 (4.4)   & 4.63 (0.13) & 1.9 (0.2) &  3.65 (0.18)	& ...	& ... \\
ON2N		& 37.4 (3.8)   & 4.71 (0.13) & 3.2 (0.3) &  3.78 (0.10)	& ...	& ... \\
S106            & 15.1 (1.7)   & 4.70 (0.17) & ...	 &  ...		& ...	& ... \\
W75N		& 76.3 (7.7)   & 4.60 (0.12) & 6.8 (0.7) &  4.15 (0.15)	& ...	& ... \\
DR21S		& 75.5 (7.6)   & 5.66 (0.15) & 7.0 (0.7) &  4.94 (0.15)	& ...	& ... \\
W75(OH)		& 91.6 (9.2)   & 5.48 (0.12) & 4.0 (0.4) &  5.44 (0.18)	& ...	& ... \\
G97.53+3.19	& 11.8 (1.3)   & 6.76 (0.31) & ...	 &  ...		& ...	& ... \\
BFS11-B		& 7.8 (1.2)    & 3.14 (0.18) & ...	 &  ...		& ...	& ... \\
CepA		& 20.2 (2.2)   & 4.07 (0.16) & ...	 &  ...		& ...	& ... \\
NGC7538		& 72.5 (7.4)   & 5.65 (0.12) & 5.2 (0.6) &  4.39 (0.16)	& ...	& ... \\
S157		& 20.4 (2.2)   & 3.51 (0.13) & ...	 &  ...		& ...	& ... \\ 
\enddata
\tablenotetext{a}{Peak position.}
\end{deluxetable}

\begin{deluxetable}{lccccc}
\tablecolumns{6}
\footnotesize
\tablecaption{Observed Properties\label{tab6}}
\tablewidth{0pt} 
\tablehead{
\colhead{Source}                  &
\colhead{Centroid}		  &
\colhead{\rcs}                    &
\colhead{\ar \tablenotemark{a} }			  &
\colhead{PA}			  &
\colhead{$R_{10}$\tablenotemark{b}}		  \\
\colhead{}                        &
\colhead{(\as , \as )}            &  
\colhead{(pc)}			  &
\colhead{}			  &
\colhead{(\degree )}	  	  &
\colhead{(pc)}			  		  
}
\startdata 
G121.30+0.66     & ($-$10,0)    & 0.10 (0.01)  & 1.5 & 55	& 0.11 (0.01) \\
G123.07-6.31     & ($-$10,0)    & 0.14 (0.01)  & 1.7 & 110	& 0.16 (0.03) \\
W3(OH)		 & (0,+10)      & 0.18 (0.01)  & 1.4 & 60	& 0.5 (0.01) \\
G135.28+2.80	 & (0,+10)      & 0.10 (0.09)  & 1.5 & 50	& U	\\
S231          	 & (0,$-$10)    & 0.17 (0.01)  & 1.2 & 135 	& 0.24 (0.03) \\
S235		 & (0,0) 	& 0.15 (0.01)  & M   & ... 	& 0.21 (0.01) \\
		 & (0,-70)	& ...	       & M   & ... 	& ... \\
S241		 & (0,+10)      & 0.23 (0.05)  & 1.7 & 90 	& U \\
S252A		 & ($-$10,+10)  & 0.10 (0.01)  & 1.2 & 135 	& U \\
		 & (+60,$-$60)	& ...	       & M   & ... 	& ... \\
S255		 & (0,0)	& ...	       & M   & ... 	& M \\
RCW142           & (0,0)        & 0.14 (0.01)  & 1.2 & 120 	& E \\
W28A2(1)         & ($-$10,0)    & 0.15 (0.04)  & 1.1 & 125 	& E \\
M8E              & (0,0)        & 0.14 (0.01)  & 1.3 & 115 	& 0.18 (0.02)\\
G9.62+0.10	 & ($-$10,+10)  & 0.33 (0.01)  & 1.3 & 35 	& 0.56 (0.05)\\
G8.67-0.36       & (0,+10)      & 0.26 (0.01)  & 1.2 & 35 	& 0.43 (0.04)\\
W31		 & ($-$10,+10)  & 0.67 (0.04)  & 1.6 & 0 	& 1.37 (0.11)\\
G10.6-0.4        & (0,+10)      & 0.41 (0.01)  & 1.0 & 45 	& E \\
G12.42+0.50	 & (0,0)	& ...	       & M   & ... 	& M \\
		 & (+10,+40)	& ...		& M  & ... 	& ... \\
G12.89+0.49      & (0,0)        & 0.19 (0.01)  & 1.3 & 115 	& 0.24 (0.04)\\
G12.2-0.1        & (0,+10)      & 0.65 (0.08)  & 1.2 & 25 	& 0.97 (0.20)\\
W33cont          & (+10,+10)    & 0.75 (0.02)  & 1.0 & ... 	& E \\
G13.87+0.28      & (0,0)        & 0.33 (0.03)  & 1.2 & 120 	& U \\
W33A		 & (+10,+10)    & 0.26 (0.01)  & 1.0 & ... 	& 0.35 (0.04)\\
G14.33-0.64	 & (0,0)        & 0.17 (0.01)  & 1.1 & 140 	& 0.29 (0.02)\\
G19.61-0.23      & (0,0)        & 0.20 (0.02)  & 1.2 & 140 	& 0.31 (0.06)\\
G20.08-0.13	 & (+10,0)	& 0.15 (0.01)  & 1.1 & 150 	& 0.19 (0.04)\\
G23.95+0.16      & (+10,0)      & 0.45 (0.03)  & 1.3 & 55 	& U \\
G24.49-0.04      & ($-$10,+10)  & 0.17 (0.01)  & 1.0 & ... 	& U \\
W42		 & ($-$10,0)    & 0.49 (0.04)  & 1.0 & ... 	& 0.64 (0.11)\\
G28.86+0.07	 & (0,+10) 	& 0.47 (0.02)  & 1.3 & 25 	& U \\
W43S		 & (0,+10)      & 0.46 (0.03)  & 1.4 & 160 	& E \\
G31.41+0.31      & (0,0)	& 0.36 (0.02)  & 1.1 & 90 	& 0.56 (0.10) \\
\enddata
\tablenotetext{a}{M = Multiple cores resulting in ambiguity.}
\tablenotetext{b}{E = Contour extended beyond map boundary, U = unresolved.}
\end{deluxetable}

\begin{deluxetable}{lccccc}
\tablecolumns{6}
\footnotesize
\tablecaption{Observed Properties Cont.\label{tab7}}
\tablewidth{0pt} 
\tablehead{
\colhead{Source}                  &
\colhead{Centroid}		  &
\colhead{\rcs}                    &
\colhead{\ar \tablenotemark{a} }			  &
\colhead{PA}			  &
\colhead{$R_{10}$ \tablenotemark{b}}		 \\
\colhead{}                        &
\colhead{(\as , \as )}            &  
\colhead{(pc)}			  &
\colhead{}			  &
\colhead{(\degree )}	 	  &
\colhead{(pc)}		
}
\startdata 
W43Main3	 & (0,0)        & 0.52 (0.05)  & 1.5 & 60 	& 1.00 (0.08)\\
G31.44-0.26      & (0,0)        & 0.52 (0.03)  & 1.0 & ... 	& 0.54 (0.13) \\
G32.05+0.06	 & (0,0)        & 0.48 (0.02)  & 1.1 & 40 	& U \\
G32.80+0.20A/B   & ($-$10,0)    & 0.96 (0.06)  & 1.1 & 55 	& 1.18 (0.17) \\
W44		 & (0,0)        & 0.37 (0.01)  & 1.2 & 45 	& E \\
S76E		 & (0,0)        & 0.20 (0.01)  & 1.6 & 130 	& 0.37 (0.01) \\
G35.58-0.03      & (0,0)        & 0.20 (0.02)  & 1.1 & 140 	& 0.21 (0.06) \\
G35.20-0.74      & (0,$-$10)    & 0.30 (0.02)  & 1.4 & 35 	& 0.45 (0.03) \\
W49N 4 \kms	 & ($-$10,0)    & 1.41 (0.04)  & ... & ... 	& E \\
W49N 12\kms	 & ($-$20,0)	& ...		& M	& ... 	& M \\
		 & (+10,+20)	& ...		& M	& ... 	& ...\\
W49S		 & (0,0)        & 1.53 (0.13)  & M   & ... 	& M \\
		 & ($-$20,+60)	& ...		& M	& ... 	& ... \\
OH43.80-0.13	 & (0,$-$10)    & 0.11 (0.01)  & 1.4 & 125 	& 0.15 (0.03) \\
G45.07+0.13      & (0,0)	& 0.48 (0.02)  & ... & ... 	& 0.70 (0.11)\\
G48.61+0.02      & (0,0)        & 0.54 (0.06)  & 1.3 & 145 	& E \\
W51W		 & ($-$20,0)    & 0.64 (0.08)  & 1.6 & 110 	& 0.82 (0.19) \\
W51M		 & (0,0)	& 0.50 (0.01)  & M   & ... 	& M \\
		 & ($-$70,+40)	& ...		& M	& ... 	& ... \\
G59.78+0.06	 & ($-$10,$+$20) & 0.18 (0.01)  & M   & ... 	& U\\
S87		 & (0,0)        & ...	       & M   & ... 	& M \\
		 & (+10,+60)	& ...		& M	& ... 	& ... \\
S88B		 & (+20,0)	& 0.16 (0.01)  & 1.0 & ... 	& 0.17 (0.02) \\
K3-50		 & (0,+10)      & 0.71 (0.05)  & 1.3 & 50 	& 0.83 (0.12) \\
ON1		 & (0,0)	& 0.43 (0.03)  & 1.0 & .. 	& 0.43 (0.09) \\
ON2S		 & ($-$10,$-$10)& 0.61 (0.02)  & 1.8 & 55 	& 0.87 (0.05) \\
ON2N		 & (0,0)        & 0.41 (0.02)  & 1.2 & 40 	& 0.64 (0.05) \\
S106             & ($+$10,0)    & 0.37 (0.06)  & ... & ... 	& U \\
W75N		 & (0,0)        & 0.27 (0.01)  & 1.5 & 70 	& 0.63 (0.02) \\
DR21S		 & (0,0)        & 0.27 (0.01)  & M   & ...  	& M \\
		 & ($-$60,0)	& ...		& M	& ... 	& ... \\
W75(OH)		 & (0,0) 	& 0.29 (0.01)  & 1.5 & 60 	& M \\
G97.53+3.19	 & ($+$10,0)	& ...	       & M   & ... 	& M \\
		 & (0,$-$20)	& ...		& M	& ... 	& ... \\
BFS11-B		 & (0,0)	& 0.12 (0.03)  & ... & ... 	& U \\
CepA		 & ($-$10,$-$10) & ...	       & M   & ... 	& M \\
		 & (+10,+10)	& ...		& M	& ... 	& ... \\
NGC7538		 & (0,0)	& 0.32 (0.01)  & M   & ... 	& M \\
		 & (0,+80)	& ...		& M	& ... 	& ... \\
S157		 & (0,+10)      & 0.19 (0.01)  & 1.0 & ... 	& 0.19 (0.03) \\ 
\enddata
\tablenotetext{a}{M = Multiple cores resulting in ambiguity.}
\tablenotetext{b}{E = Contour extended beyond map boundary, U = unresolved.}
\end{deluxetable}

\begin{deluxetable}{lccccccc}
\tablecolumns{8}
\footnotesize
\tablecaption{Derived Properties\label{tab8}}
\tablewidth{0pt} 
\tablehead{
\colhead{Source}                  &
\colhead{$p$\tablenotemark{a}}			  &
\colhead{$M_{vir}$(\rcs )}            &
\colhead{$M_{vir}(R_n)$}            &
\colhead{$\Sigma$}     	          &
\colhead{$\log X$(CS)}                     &
\colhead{$L$(CS5--4)}		  &
\colhead{$L_{bol}/M_{vir}$}      \\
\colhead{}                        &  
\colhead{}			  &
\colhead{(\msun )}                &
\colhead{(\msun )}                &
\colhead{(g cm$^{-2}$)}		  &
\colhead{}  			  &
\colhead{($10^{-2}$ \lsun )}       &
\colhead{(\lsun $/$\msun )}
}
\startdata 
G121.30+0.66     & 1.25 & 320 (80) 	& 1870	& 2.16 (0.59)	& $-$10.53 (0.13)	& 0.10 (0.03)	& 3 \\
G123.07-6.31     & 1.75 & 500 (200)	& 1640	& 1.72 (0.73)	& ...			& 0.28 (0.09)	& 12 \\
W3(OH)		 & 1.50 & 1020 (130) 	& 3550	& 2.08 (0.29)	& $-$9.57 (0.36)	& 1.15 (0.27)	& 93 \\
G135.28+2.80	 & ...	& 210 (110)	& ...	& 0.12 (0.09)   & $-$8.86 (0.42)	& 0.60 (0.34	& 269 \\
S231          	 & 1.50 & 180 (50)	& 490	& 0.40 (0.12)   & $-$8.74 (0.41)	& 0.40 (0.13)	& 73 \\
S235		 & ...	& 100 (40)	& ...	& 0.29 (0.11)   & $-$9.43 (0.12)	& 0.30 (0.08)	& 98 \\
S241		 & ...	& 140 (60)	& ...	& 0.18 (0.11)   & ...			& 0.30 (0.16)	& 91 \\
S252A		 & 1.75 & 140 (90)	& 350	& 0.99 (0.67)   & ...			& 0.09 (0.03)	& 45 \\
RCW142           & 2.25 & 370 (290)	& 610	& 1.23 (0.95)   & $-$8.28 (1.37)	& 1.19 (0.27)	& 153 \\
W28A2(1)         & 2.25	& 450 (340)	& 1280	& 1.29 (0.97)   & $-$7.78 (3.75)	& 2.85 (0.65)	& 450 \\
M8E              & 1.75 & 100 (30) 	& 200	& 0.37 (0.11)   & $-$8.14 (0.89)	& 0.29 (0.07)	& 166 \\
G9.62+0.10	 & 2.00 & 2230 (870)	& 3930	& 1.37 (0.54)   & $-$8.50 (0.89)	& 3.68 (0.87)	& 157 \\
G8.67-0.36       & 2.00 & 860 (340)	& 1890	& 0.82 (0.33)   & ...			& 1.97 (0.47)	& 152 \\
W31		 & ...	& 7300 (1670)	& ...	& 1.09 (0.28)   & $-$8.44 (0.64)	& 15.9 (4.1)	& ... \\
G10.6-0.4        & 2.50 & 2750 (460)	& ...	& 1.10 (0.19)   & $-$7.90 (2.44)	& 17.2 (3.9)	& 334 \\
G12.89+0.49      & 2.00 & 340 (130)	& 470	& 0.63 (0.25)   & $-$9.08 (0.29)	& 0.71 (0.19)	& 115 \\
G12.2-0.1        & ...	& 4810 (1480)	& ...	& 0.77 (0.30)   & $-$8.83 (0.59)	& 11.3 (3.9)	& 114 \\
W33cont          & ...	& 2950 (520)    & ...	& 0.35 (0.06)   & $-$8.27 (0.69)	& 22.7 (5.7)	& ... \\
G13.87+0.28      & 1.75 & 310 (120)	& 350	& 0.19 (0.08)   & $-$8.60 (0.46)	& 0.92 (0.32)	& 419 \\
W33A		 & 1.50 & 454 (130)	& 1260	& 0.44 (0.13)   & ...			& 1.33 (0.34)	& 220 \\
G14.33-0.64	 & 2.00 & 160 (60)	& 450	& 0.37 (0.14)   & $-$8.48 (0.39)	& 0.85 (0.20)	& 621 \\
G19.61-0.23      & ...	& 1270 (380)	& ...	& 2.08 (0.73)   & $-$9.19 (0.39)	& 1.55 (0.48)	& 141 \\
G20.08-0.13	 & ...	& 1610 (460)	& ...	& 4.56 (1.42)   & $-$9.83 (0.45)	& 0.51 (0.13)	& ... \\
G23.95+0.16      & 1.50 & 430 (150)	& 270	& 0.14 (0.05)   & ...			& 1.72 (0.55)	& 443 \\
G24.49-0.04      & 2.25 & 300 (130)	& 450	& 0.67 (0.31)   & $-$8.98 (0.67)	& 0.38 (0.10)	& 164 \\
W42		 & ...	& 2160 (480)	& ...	& 0.60 (0.16)   & $-$8.78 (0.39)	& 5.66 (1.59)	& ... \\
G28.86+0.07	 & ...	& 710 (200)	& ...	& 0.21 (0.06)   & $-$8.71 (0.46)	& 2.33 (0.55)	& ... \\
W43S		 & 2.50 & 1080 (230)	& ...	& 0.34 (0.08)   & $-$8.97 (0.45)	& 7.31 (2.01)	& 1480 \\
G31.41+0.31      & 2.25	& 1040 (930)	& 2090	& 0.53 (0.48)   & $-$9.16 (0.39)	& 4.62 (1.17)	& 221 \\
\enddata
\tablenotetext{a}{$p = -\log n/\log r$ from Mueller et al. 2002}
\end{deluxetable}

\begin{deluxetable}{lccccccc}
\tablecolumns{8}
\footnotesize
\tablecaption{Derived Properties Cont.\label{tab9}}
\tablewidth{0pt} 
\tablehead{
\colhead{Source}                  &
\colhead{$p$\tablenotemark{a}}			  &
\colhead{$M_{vir}$(\rcs )}            &
\colhead{$M_{vir}(R_n$)}            &
\colhead{$\Sigma$}     	          &
\colhead{$\log X$(CS)}                     &
\colhead{$L$(CS5--4)}		  &
\colhead{$L_{bol}/M_{vir}$}      \\
\colhead{}                        &
\colhead{}			  & 
\colhead{(\msun )}                &
\colhead{(\msun )}                &
\colhead{(g cm$^{-2}$)}		  &
\colhead{}  			  &
\colhead{($10^{-2}$ \lsun )}      &
\colhead{(\lsun $/$\msun )}
}
\startdata 
W43Main3	 & ...	& 3610 (960)	& ...	& 0.89 (0.29)  & ...		& 4.84 (1.88)	& ... \\
G31.44-0.26      & ...	& 1120 (490)	& ...	& 0.28 (0.13)  & $-$9.12 (0.38)	& 4.56 (1.18)	& ... \\
G32.05+0.06	 & ...	& 1470 (790)	& ...	& 0.43 (0.23)  & $-$8.81 (0.63)	& 2.09 (0.52)	& ... \\
G32.80+0.20A/B   & ...	& 3800 (2280)	& ...	& 0.28 (0.17)  & $-$8.68 (0.62)	& 14.8 (4.07)	& ... \\
W44		 & ...	& 1400 (600)	& ...	& 0.68 (0.29)  & $-$8.92 (0.67)	& 5.95 (1.45)	& 214 \\
S76E		 & 1.50 & 240 (20)	& ...	& 0.41 (0.04)  & $-$8.39 (0.58)	& 0.90 (0.21)	& 118 \\
G35.58-0.03      & ...	& 1280 (500)	& ...	& 2.15 (0.93)  & $-$9.40 (0.49)	& 0.56 (0.19)	& 33 \\
G35.20-0.74      & ...	& 3200 (1190)	& ...	& 2.37 (0.90)  & $-$9.35 (0.69)	& 1.22 (0.35)	& ... \\
W49N 4\kms	 & ...	& 14570 (5960) & ...	& 0.94 (0.39)  & $-$8.56 (0.62)	& 52.8 (13.4)	& ... \\
W49S		 & ...	& 13030 (4140) & ...	& 0.37 (0.13)  & $-$8.61 (0.32)	& 17.9 (7.2)	& ... \\
OH43.80-0.13	 & ...	& 270 (100)	& ...	& 1.59 (0.66)  & $-$9.14 (0.34)	& 0.32 (0.08)	& ... \\
G45.07+0.13      & ...	& 2690 (580)	& ...	& 0.77 (0.18)  & $-$9.02 (0.46)	& 7.14 (1.73)	& 446 \\
G48.61+0.02      & ...	& 440 (200)	& ...	& 0.10 (0.06)  & $-$8.37 (0.51)	& 3.50 (1.13)	& 2290 \\
W51W		 & ...	& 1100 (480)	& ...	& 0.18 (0.09)  & $-$9.32 (0.32)	& 3.81 (1.85)	& ... \\
W51M		 & ...	& 5930 (980)	& ...	& 1.61 (0.28)  & $-$8.40 (1.45)	& 28.7 (6.6)	& 472 \\
G59.78+0.06	 & ...	& 30 (20)	& ...	& 0.07 (0.05)  & $-$8.13 (0.56)	& 0.25 (0.09)	& ... \\
S88B		 & 1.25 & 160 (40)	& 220	& 0.41 (0.10)  & $-$9.26 (0.18)	& 0.26 (0.07)	& 562 \\
K3-50		 & ...	& 6130 (2140)	& ...	& 0.81 (0.30)  & $-$10.15 (1.19)& 5.92 (1.91)	& 343 \\
ON1		 & 1.75 & 1320 (370)	& 2230	& 0.48 (0.15)  & ...		& 1.87 (0.58)	& 114 \\
ON2S		 & 1.75 & 1220 (300)	& 700	& 0.22 (0.06)  & $-$9.33 (0.30)	& 6.02 (1.60)	& 302 \\
ON2N		 & ...	& 870 (210)	& ...	& 0.35 (0.09)  & ...		& 3.04 (0.88)	& ... \\
S106             & ...	& 720 (180)	& ...	& 0.35 (0.11)  & ...		& 0.88 (0.39)	& 692 \\
W75N		 & ...	& 700 (180)	& ...	& 0.63 (0.13)  & $-$9.59 (0.41)	& 2.41 (0.60)	& ... \\
DR21S		 & ...	& 990 (220)	& ...	& 0.90 (0.22)  & $-$9.79 (0.32)	& 1.78 (0.48)	& 506 \\
W75(OH)		 & ...	& 1260 (250)	& ...	& 1.03 (0.21)  & $-$9.41 (0.34)	& 2.30 (0.56)	& 40 \\
BFS11-B		 & ...	& 110 (50)	& ...	& 0.48 (0.31)  & ...		& 0.07 (0.05)	& 66 \\
NGC7538		 & ...	& 920 (190)	& ...	& 0.60 (0.12)  & $-$9.39 (0.09)	& 2.83 (0.66)	& 206 \\
S157		 & 0.75 & 200 (20)	& 450	& 0.39 (0.06)  & $-$9.91 (1.05)	& 0.35 (0.10) 	& 141 \\ 
\enddata
\tablenotetext{a}{$p = -\log n/\log r$ from Mueller et al. 2002}
\end{deluxetable}

\begin{deluxetable}{lcccccccc}
\tablecolumns{9}
\footnotesize
\tablecaption{Statistical Summary\label{tab10}}
\tablewidth{0pt} 
\tablehead{
\colhead{Source}		&
\colhead{Sample\tablenotemark{c}} 		&
\colhead{N}			&
\colhead{Mean}			& 
\colhead{Standard}		&
\colhead{Mean}			&
\colhead{Median}		&
\colhead{Skewness\tablenotemark{c}}		&
\colhead{Units}			\\
\colhead{Property}		&
\colhead{}			&
\colhead{}			&
\colhead{}			&
\colhead{Deviation}		&
\colhead{Deviation\tablenotemark{b}}		&
\colhead{}			&
\colhead{}			&
\colhead{}
}
\startdata 
D			  & Total & 63	& 5.3	& 3.7	& 3.0	& 4.0	& 1.1	& kpc \\
D$_g$			  & Total & 63	& 7.2	& 2.6	& 2.2	& 6.8	& 0.5   & kpc \\
I$_{peak}$(T$_R^*$, \cs ) & Total & 63	& 47.2	& 44.7	& 30.1	& 31.8	& 2.4   & K \kms \\
			  & NoRC  & 12  & 25.9  & 23.1  & 14.8  & 19.7  & 2.3   & K \kms \\
			& UC\HII\ & 32  & 50.4  & 41.0  & 29.1  & 35.7  & 2.1	& K \kms \\
		   & C\HII /\HII\ & 19  & 54.6  & 58.0  & 37.7  & 32.0  & 2.3   & K \kms \\
(S/N)$_{peak}$		  & Total & 63	& 50	& 40	& 27	& 40	& 2.4   & \\
I$_{peak}$(T$_R^*$, \css )& Total & 49	& 6.5	& 7.5	& 5.2	& 4.0	& 1.9	& K \kms \\
\rcs			  & Total & 57	& 0.37	& 0.26	& 0.19	& 0.32	& 2.0   & pc \\
			  & NoRC  & 12  & 0.28  & 0.14  & 0.12  & 0.25  & 0.6   & pc \\
			& UC\HII\ & 27  & 0.38  & 0.32  & 0.22  & 0.27  & 2.2   & pc \\
		   & C\HII /\HII\ & 18  & 0.43  & 0.22  & 0.17  & 0.41  & 0.7   & pc \\
$R_{10}$	 	  & Total & 33	& 0.50  & 0.32  & 0.26  & 0.43  & 0.9   & pc \\
\dv			  & Total & 63	& 5.6	& 2.2	& 1.8	& 5.1	& 0.7   & \kms \\
\dvs			  & Total & 51	& 5.0	& 2.0	& 1.7	& 4.9	& 0.4   & \kms \\
\dvc			  & Total & 9	& 5.7	& 2.0	& 1.5	& 5.3	& 0.3	& \kms \\
\ar			  & Total & 47	& 1.27	& 0.22	& 0.18	& 1.20	& 0.6   & \\
M$_{vir}$(\rcs ) 	  & Total & 57	& 1810	& 2810	& 1750	& 920	& 3.1   & \msun \\
			  & NoRC  & 12  & 760   & 1000  & 678   & 330   & 2.5   & \msun \\
			& UC\HII\ & 27  & 2170  & 3650  & 2200  & 990   & 2.8   & \msun \\
		   & C\HII /\HII\ & 18  & 1960  & 1990  & 1650  & 1160  & 1.1   & \msun \\
$\log$ M$_{vir}$(\rcs )   & Total & 57  & 2.90  & 0.57  & 0.46  & 2.97  & 0.04  & $\log$ \msun \\
M$_{vir}$(R$_n$) 	  & Total & 21	& 1180	& 1080	& 870	& 610	& 1.3   & \msun \\
$\Sigma$(\rcs )  	  & Total & 57	& 0.82	& 0.78	& 0.55	& 0.60	& 2.4   & g cm$^{-2}$ \\
$f_v(p = 0)$		  & Total & 42	& 0.47	& 0.72	& 0.50	& 0.14	& 2.4	& \\
X(CS)			  & Total & 46	& 3.0	& 5.9	& 3.1	& 1.1	& 4.4	& $10^{-9}$ \\
$\log$ X(CS)		  & Total & 46  & $-$8.93 & 0.62& 0.49  & $-$8.94& $-$0.1 & \\  
\lcs			  & Total & 57	& 5.0	& 8.8	& 5.2	& 1.9	& 3.6   & $10^{-2}$ \lsun \\
$\log$ \lcs		  & Total & 57    & $-$1.8  & 0.7   & 0.5   & $-$1.7 & 0.01 & $\log$ \lsun \\
L$_{bol}$/M$_{vir}$(\rcs )& Total & 40  & 310	& 420	& 250	& 160	& 3.4   & \lsun /\msun \\
			  & NoRC  &  9  & 120   & 90    & 70    & 90    & 0.7   & \lsun /\msun \\
			& UC\HII\ & 15  & 300   & 220   & 200   & 170   & 0.6   & \lsun /\msun \\
		   & C\HII /\HII\ & 16  & 440   & 600   & 370   & 260   & 2.5   & \lsun /\msun \\
$\log$ L$_{bol}$/M$_{vir}$(\rcs ) & Total & 40 & 2.24 & 0.52 & 0.38 & 2.22 & $-$0.8 & $\log$ \lsun /\msun \\  
$\mean{\bar P/k}$	  & Total & 57	& 5.4	& 12.6	& 6.3	& 1.5	& 5.4	& $10^8$ K \cmv \\
$\log$ $\mean{\bar P/k}$  & Total & 57    & 8.14  & 0.76  & 0.61  & 8.18  & $-$0.1 & $\log$ K \cmv \\
\enddata
\tablenotetext{a}{Distribution sample.  Total = complete sample.  NoRC = no known radio continuum.
UC\HII\ = contains UC\HII\ region.  C\HII /\HII\ = assocaited with C\HII\ or \HII\ region.}
\tablenotetext{b}{Mean Deviation = $\frac{1}{N}\sum_i |x_i - <x>|$}
\tablenotetext{c}{Skewness = $\sum_i \frac{(x_i - <x>)^3}{(N - 1)\sigma^3}$}
\end{deluxetable}

\end{document}